\documentclass[twocolumn]{aastex701}
\usepackage{amsmath}   %

\newcommand{\Lcdm}{$\Lambda$CDM}
\newcommand{\dolphot}{\texttt{DOLPHOT}}
\newcommand{\euclid}{\textit{Euclid}}
\newcommand{\rst}{\textit{Roman}}
\newcommand{\hst}{\textit{HST}}
\newcommand{\jwst}{\textit{JWST}}

\shorttitle{A \dolphot{}/\euclid{} Study of AM\nobreakdash-1}
\shortauthors{Weisz et al.}
\journalinfo{Accepted to PASP}

\begin{document}

\title{A \dolphot{}/\euclid{} Study of the Resolved, Faint Outer Milky Way Globular
Cluster AM\nobreakdash-1}

\author[0000-0002-6442-6030]{Daniel~R.\ Weisz}\affil{Department of Astronomy, University of California, Berkeley, CA 94720-3411, USA}\affil{Miller Institute for Basic Research, University of California Berkeley, Berkeley, CA, 94720, USA}\email[show]{dan.weisz@berkeley.edu}\correspondingauthor{Daniel~R.\ Weisz}

\author[0000-0002-1445-4877]{Alessandro~Savino}\affil{Department of Astronomy, University of California, Berkeley, CA 94720-3411, USA}\email{astrosavino@gmail.com}

\author[0000-0001-8416-4093]{Andrew~E.\ Dolphin}\affil{Raytheon, 1151 E. Hermans Road, Tucson, AZ 85756, USA}\affil{Steward Observatory, University of Arizona, 933 N. Cherry Avenue, Tucson, AZ 85719, USA}\email{adolphin@rtx.com}

\author[0000-0001-7531-9815]{Meredith~J.~Durbin}
\affiliation{Department of Astronomy, University of Washington, Box 351580, Seattle, WA 98195-1580, USA}
\affiliation{Department of Astronomy, University of California, Berkeley, CA 94720-3411, USA}
\email{mdurbin@uw.edu}

\author[0000-0002-5865-0220]{Abigail~J.~Lee}\altaffiliation{NASA Hubble Fellow}\affil{Department of Astronomy, University of California, Berkeley, CA 94720, USA}\affil{Miller Institute for Basic Research, University of California Berkeley, Berkeley, CA, 94720, USA}\email{abby.lee@berkeley.edu}

\author[0000-0003-0605-8732]{Evan~D.\ Skillman}\affil{University of Minnesota, Minnesota Institute for Astrophysics, School of Physics and Astronomy, 116 Church Street SE, Minneapolis, MN 55455, USA}\email{skill001@umn.edu}

\begin{abstract}
We present a resolved star study of AM\nobreakdash-1, a distant ($D \simeq 120$~kpc), faint
($M_V \sim -5.1$) Milky Way (MW) globular cluster, based on \euclid{} Quick Data
Release~1 (Q1) imaging. To perform photometry of the \euclid{} VIS and
NISP imaging, we develop a \euclid-specific module for the crowded field stellar
photometry package \dolphot{}.
We construct and present empirical point spread function (PSF) models for VIS and
NISP, along with a photometric calibration, entirely from the public Q1 images.
Our photometry reaches $5\sigma$ point source depths of
$I_E, Y_E, J_E, H_E = 27.1, 25.3, 25.5, 25.3$, $\sim$1--1.5~mag deeper than the
public Q1 catalog, as expected for dedicated resolved star PSF photometry. We resolve the cluster to $\sim$1~mag below its main sequence
turnoff in all bands. The \euclid{} photometry agrees with color--magnitude diagram (CMD) modeling of
archival \hst/WFPC2 imaging and with the 2MASS-anchored VISTA Hemisphere Survey to
$\lesssim0.02$~mag in $I_E$, $J_E$, and $H_E$, with a $\sim$0.07~mag offset in $Y_E$.
We find that AM\nobreakdash-1 hosts an old ($\sim$10.6~Gyr),
intermediate-metallicity ([Fe/H]~$\sim-1$) population, with a present-day
stellar mass of $\sim$$2\times10^4\,M_\odot$ and $M/L_V \simeq 2.2$, and has a half-light radius of $\sim$12--16~pc with a round,
low-concentration profile. These affirm earlier studies placing AM\nobreakdash-1 among the most extended, diffuse
outer-halo clusters, which dynamical work identifies as accreted.  We describe and publicly release \dolphot{} v2.1, the \euclid{} module, and its empirical PSFs, which together enable \euclid{} resolved star studies ranging from faint MW systems such
as AM\nobreakdash-1 to more luminous galaxies throughout the Local Volume. We anticipate further
refinements, including spatially varying PSFs, with \euclid{} Data Release~1.

\end{abstract}

\keywords{Globular star clusters (656) --- Milky Way stellar halo (1060) --- Stellar photometry (1620) --- Hertzsprung Russell diagram (725) --- Space telescopes (1547)}

\section{Introduction}
\label{sec:intro}

Faint dwarf galaxies and
low-mass star clusters that populate the halo of the Milky Way (MW) are among
our most sensitive probes of dark matter, reionization, and the hierarchical
assembly of galaxies, and they have long anchored our understanding of \Lcdm{} on small scales \citep[e.g.,][]{Bullock:2017, Simon:2019}. Realizing the scientific potential of these low-luminosity systems
requires finding them and characterizing their stellar populations, both of which are limited by the depth, area, and angular resolution of the available imaging.

The Sloan Digital Sky Survey (SDSS) opened this faint regime, with the
discovery of ultra-faint dwarf
galaxies \citep[UFDs; e.g.,][]{Willman:2005a, Willman:2005b,
Zucker:2006a, Zucker:2006b, Belokurov:2007}. Since that time, deeper imaging from programs such as the Dark Energy Survey \citep[DES;][]{DES:2016}, Pan-STARRS \citep{Chambers:2016}, the DECam Local Volume Exploration survey \citep[DELVE;][]{Drlica-Wagner:2021}, and the Ultraviolet Near-Infrared Optical Northern Survey \citep[UNIONS;][]{Gwyn:2025} has roughly doubled the satellite census again while pushing discovery to lower surface brightness and larger distances \citep[e.g.,][]{Bechtol:2015, Koposov:2015, Drlica-Wagner:2015, Laevens:2015a, Mau:2020, Cerny:2023c, Smith:2024}.  The Vera C. Rubin Observatory and its Legacy
Survey of Space and Time (LSST) are expected to further expand the census of faint systems around the MW
\citep[e.g.,][]{Ivezic:2019, Mutlu:2021, Tsiane:2025, Shread:2026, Tan:2026}.

Wide-area, space-based imaging offers a further advance by providing the depth, angular resolution, and point spread function (PSF) stability needed to resolve
these systems into individual stars and construct clean, deep color--magnitude diagrams (CMDs).
The Nancy Grace Roman Space Telescope (\rst), \euclid, and the Chinese Space Station
Telescope (\textit{CSST}) will each image large fractions of the sky with image quality that approaches what has only been available to date
over the small fields of the Hubble Space Telescope (\hst) and the James Webb
Space Telescope (\jwst) \citep[e.g.,][]{Akeson:2019, Mellier:2025, Gong:2026}.
This combination is ideally suited to discovering and
characterizing low-luminosity systems around the MW and, more broadly,
 resolved stellar systems of all luminosities throughout the Local Volume ($D < 10$ Mpc). 

Here, we demonstrate the potential of \euclid{} for the study of resolved stellar
systems by focusing on the low-luminosity MW globular cluster (GC) AM\nobreakdash-1. AM\nobreakdash-1 is the
sole known Galactic GC contained in the \euclid{} Quick Data Release~1 (Q1), which makes it
the natural testbed for resolved star photometry with the publicly available
data.
First identified on photographic survey plates by \citet{Madore:1979},
AM\nobreakdash-1 is among the most distant MW GCs ($D\sim$120 kpc; e.g., \citealt{Suntzeff:1985, Dotter:2008b}).
This large distance, along with its sparse stellar population, has
long made AM\nobreakdash-1 a challenging target. We summarize the properties of AM\nobreakdash-1 in Table~\ref{tab:am1}.

The most detailed view of AM\nobreakdash-1 to date comes from deep \hst/WFPC2 (Wide Field Planetary Camera 2) imaging in F555W and F814W
by \citet{Dotter:2008b} that reaches well below the main sequence turnoff (MSTO). They found that AM\nobreakdash-1 is an old, metal-poor ($[\mathrm{Fe/H}]=-1.5$) cluster that features a prominent red horizontal branch (HB). In this respect AM\nobreakdash-1
resembles other distant, outer halo GCs such as Pal\,3, Pal\,4,
Pal\,14, and Eridanus, which also have red HBs and comparable ages
\citep{Stetson:1999, Dotter:2008b}.

To enable studies of resolved stellar systems such as AM\nobreakdash-1, we develop a \euclid-specific module for \dolphot{}\footnote{\url{http://americano.dolphinsim.com/dolphot/}},
a crowded field stellar PSF photometry package that is widely used for studies of
nearby galaxies \citep[e.g.,][]{Dolphin:2000, Dalcanton:2009, Radburn-Smith:2011,
Williams:2014, Savino:2025}. The module provides a robust,
well-tested, and publicly available route to PSF photometry of \euclid{} VIS and
NISP imaging within a framework that includes other telescope-specific modules for
\hst, \jwst, and \rst{} data \citep[e.g.,][]{Dalcanton:2012b, Weisz:2024}. The current \euclid{} module
relies only on publicly released Q1 data products and requires no
consortium-internal tools or PSF models. Early \euclid{} data, notably the Early Release Observations (ERO) of nearby
galaxies \citep{Hunt:2025} and MW GCs \citep{Massari:2025, Griggio:2026}, are already delivering high-precision resolved star science, including
empirical PSF modeling and astrometry of Galactic GCs \citep{Libralato:2024,
Libralato:2026}, multiple population diagnostics in \euclid{} colors \citep{Milone:2025},
and deep photometry of nearby galaxies to faint surface brightness \citep{Annibali:2026}.

However, these analyses have largely relied on data products internal to the Euclid
Consortium, i.e., individual calibrated frames rather than the public survey releases,
and, in most cases, on specialized photometry codes developed within that community
(e.g., \texttt{KS2}, \texttt{euclid1pass}).
A main goal of our \euclid{} Guest Investigator Program (24-EGIP24-0005, PI Savino) is the public release of this \dolphot{}
module for \euclid{}, which will enable the broader community to maximize the scientific
return of \euclid{} starting with the first major data release (DR1).

\begin{figure}[t]
\centering
\includegraphics[width=\columnwidth]{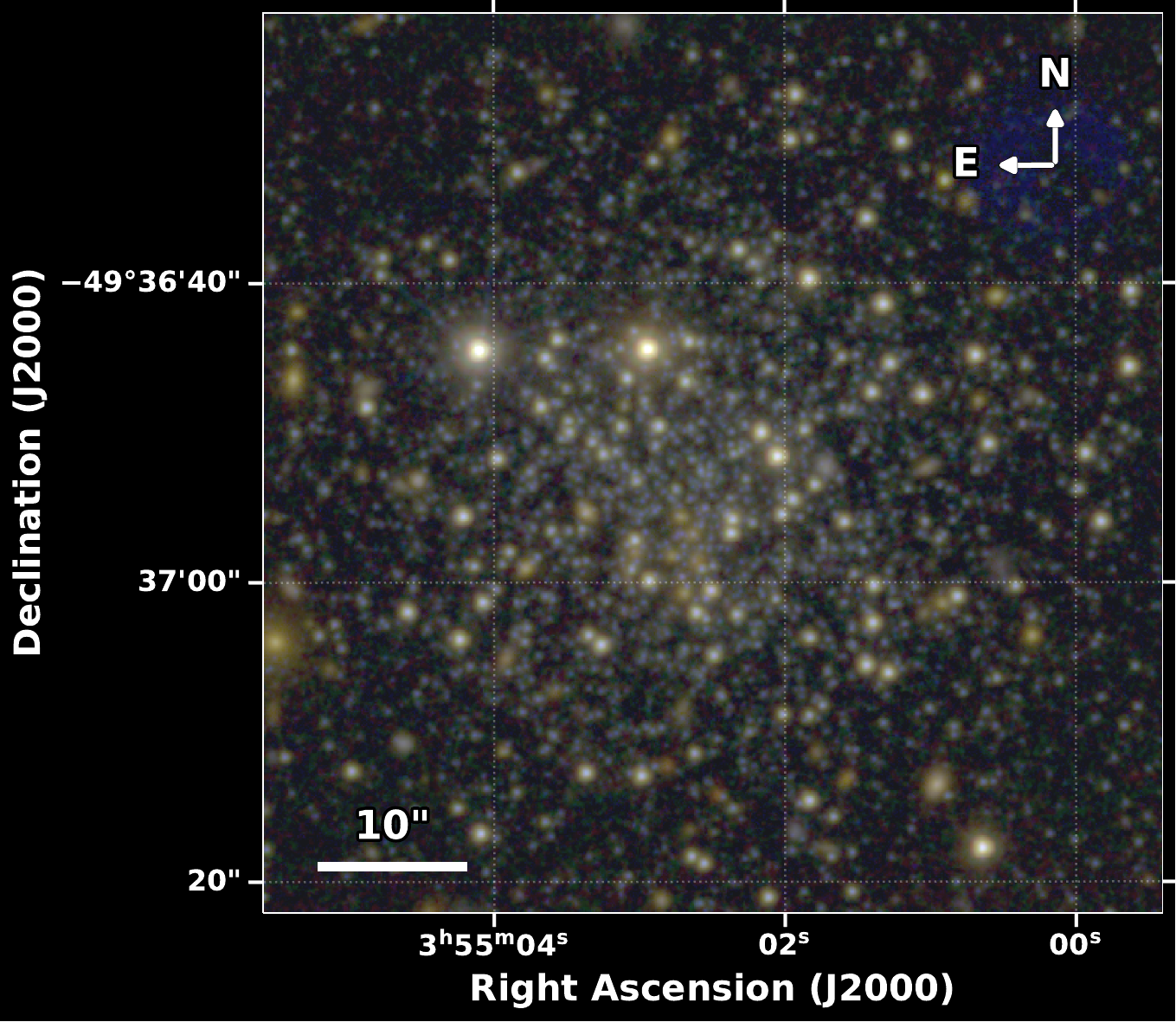}
\caption{\euclid{} color image of AM\nobreakdash-1, constructed from the Q1 VIS and NISP
mosaics ($I_E$ blue, $J_E$ green, $H_E$ red). The field is $\sim1\arcmin$ on a
side and is centered on the cluster. The
cluster is resolved into individual stars across its full extent, illustrating the
angular resolution and sensitivity that \euclid{} brings to bear on distant, sparse
stellar systems.
\label{fig:am1_color}}
\end{figure}

\begin{deluxetable}{ll}
\tabletypesize{\footnotesize}
\tablewidth{0pt}
\tablecaption{Observational and Structural Properties of AM-1\label{tab:am1}}
\tablehead{\colhead{Quantity} & \colhead{Value}}
\startdata
\cutinhead{Observational}
R.A. (J2000)                         & 03:55:02.9              \\
Decl. (J2000)                        & $-$49:36:50            \\
\euclid\ obs.\ date                  & 2024 Sep 6             \\
\euclid\ exp.\ ($I_E$)               & 1861 s                 \\
\euclid\ exp.\ ($Y_E,J_E,H_E$)       & 349 s each             \\
WFPC2 obs.\ date                     & 1999 Jun               \\
WFPC2 exp.\ (F555W, F814W)           & 16,120 s / 19,120 s\tablenotemark{a} \\
$5\sigma$ limit ($I_E,Y_E,J_E,H_E$)  & 27.1, 25.3, 25.5, 25.3 \\
50\% compl.\ ($I_E,Y_E,J_E,H_E$)     & 26.6, 25.1, 25.1, 24.9 \\
Stars in \euclid\ catalog ($<60''$)  & 4113\tablenotemark{e}  \\
CMD-selected structural candidates   & 306\tablenotemark{f}   \\
\cutinhead{Physical (WFPC2, this work)}
$(m-M)_0$                            & $20.38 \pm 0.01$       \\
$D$ (kpc)                            & $119 \pm 1$            \\
Age (Gyr)                            & $10.6 \pm 0.6$         \\
$\log(t/\mathrm{yr})$                & $10.025 \pm 0.025$     \\
$[\mathrm{Fe/H}]$                    & $-1.00 \pm 0.07$\tablenotemark{d} \\
$A_V$                                & $0.022$\tablenotemark{b} \\
$M_V$                                & $-5.1$\tablenotemark{c} \\
$L_V$ ($L_\odot$)                     & $9.4\times10^3$        \\
$M_{\rm present}$ ($M_\odot$)         & $\sim2.1\times10^4$    \\
$M/L_V$ ($M_\odot/L_\odot$)          & $\sim2.2$              \\
\cutinhead{Structural (\euclid, this work)}
$r_{h,\rm exp}$ (arcsec)           & $21.6^{+1.2}_{-1.3}$   \\
$r_{h,\rm exp}$ (pc)               & $12.5^{+0.8}_{-0.7}$   \\
$r_{h,\rm Plummer}$ (arcsec)       & $20.8^{+1.1}_{-1.2}$   \\
$r_{h,\rm Plummer}$ (pc)           & $12.1^{+0.7}_{-0.7}$   \\
$r_{h,\rm King}$ (arcsec)          & $27.0^{+1.4}_{-1.4}$   \\
$r_{h,\rm King}$ (pc)              & $15.7^{+0.8}_{-0.8}$   \\
$W_0$ (King)                         & $4.8^{+0.4}_{-0.5}$    \\
$c = \log(r_t/r_0)$                  & $0.98^{+0.09}_{-0.09}$ \\
$1-b/a$                              & $<0.12$ (90\%)         \\
\enddata
\tablenotetext{a}{\citet{Dotter:2008b}.}
\tablenotetext{b}{\citet{Schlafly:2011}.}
\tablenotetext{c}{Absolute magnitude from \citet{munoz:2018b} corrected to our adopted distance
($D = 119$~kpc) and extinction ($A_V = 0.022$).}
\tablenotetext{d}{Adopted from the CMD fit to multiple stellar models.}
\tablenotetext{e}{Sources within $60\arcsec$ passing our $SNR$ and sharpness cuts in at
least one band.} %
\tablecomments{All quantities are derived in this work unless a footnote indicates
otherwise. }%
\end{deluxetable}

This paper is organized as follows. In \S~\ref{sec:data} we summarize the
\euclid{} VIS and NISP imaging of AM\nobreakdash-1 and the supporting \hst/WFPC2 dataset. In \S~\ref{sec:psf} we
detail our process of empirical PSF model construction for VIS and NISP. In \S~\ref{sec:module} we describe the \euclid{} \dolphot{}
module, including new features of \dolphot{} v2.1, and our preprocessing of the VIS and NISP images. In \S~\ref{sec:phot} we present our
\dolphot{} photometry of AM\nobreakdash-1. In
\S~\ref{sec:discussion} we compare it with the Q1 MER catalog and ground-based near-infrared photometry, assess the stellar populations of AM\nobreakdash-1 using \euclid{} and WFPC2 CMDs, and measure the cluster's structural parameters.
Finally, we summarize
our results and describe the public release of the module in
\S~\ref{sec:summary}.

\section{Data}
\label{sec:data}

\subsection{Euclid}

\euclid{} \citep{Laureijs:2011} observed AM\nobreakdash-1 with both the VIS \citep{Cropper:2025} and NISP \citep{Jahnke:2025} in September 2024. We retrieved the data from the NASA/IPAC Infrared Science Archive (IRSA) using the Python \texttt{Irsa} interface, selecting the calibrated frames (for VIS, quadrant frames, i.e., the $\sim$2k$\times$2k region read out by one of a CCD's four amplifiers) of Observation~ID~3590, which contains AM\nobreakdash-1. The full set of exposures is listed in Table~\ref{tab_euclid_exposures}. Our analysis of AM\nobreakdash-1 uses only the three VIS quadrants and single NISP detector that cover the cluster (Table~\ref{tab_euclid_exposures}), rather than the full field of view of Observation~3590. However, we build the PSF models (\S~\ref{sec:psf}) from all detectors in all 124 Q1 observations.

The cluster was imaged at four dither pointings. Each NISP pointing consists of a single 87~s exposure in each of $Y_E$, $J_E$, and $H_E$. The nominal \euclid{} observing sequence acquires one 561~s VIS exposure at each of the four dithers, along with a 90~s VIS exposure at the first two \citep{McCracken:2026}, but the AM\nobreakdash-1 VIS data comprise two 90~s and three 561~s exposures in $I_E$. The deep exposure from the first dither is absent from Q1, and we have confirmed that it will also be absent from DR1, presumably because it did not pass the validation checks required for release (IRSA Helpdesk, private communication). The \euclid{} pipeline flags cosmic rays in the calibrated frames, and we therefore required no additional preprocessing before running \dolphot{}. Figure~\ref{fig:am1_color} shows a color image of AM\nobreakdash-1 constructed from the VIS and NISP mosaics. The cluster is cleanly resolved into individual stars across its full extent.

\subsection{Hubble Space Telescope}
\label{hst_data}

We complement the \euclid{} imaging with archival \hst/WFPC2 observations of AM\nobreakdash-1, which we use as an external check on our \euclid{} photometry (see \S~\ref{subsec:AM1_population}). The observations were obtained in June 1999 in F555W and F814W, comprising four short (180--230~s) and 14 long (900--1400~s) exposures in each band. We refer the reader to \citet{Dotter:2008b} for full details.

We retrieved the WFPC2 images from the Mikulski Archive for Space Telescopes (MAST). The cluster falls on the WF3 chip, which was dithered with integer and half-pixel offsets. We therefore cleaned cosmic rays only within groups of images sharing a common filter and an integer-pixel shift. This yields four groups of 4--5 images per filter, which we find provides adequate cosmic-ray rejection prior to photometry. As a final step, we combine any deep images taken at a common dither position to speed up the photometry.

\begin{deluxetable}{lllrr}
\tablewidth{0pt}
\tablecaption{Euclid observations of AM\nobreakdash-1, all taken September 6 2024. \label{tab_euclid_exposures}}
\tablehead{
\colhead{Image} &
\colhead{Start Time} &
\colhead{Detector} &
\colhead{Filter} &
\colhead{Exposure}}
\startdata
3590-0 & 03:35:27 & NISP DET21 & $J_E$ & 87.2448 \\
3590-0 & 03:37:41 & NISP DET21 & $H_E$ & 87.2448 \\
3590-0 & 03:39:55 & NISP DET21 & $Y_E$ & 87.2448 \\
3590-0-2 & 03:39:58 & VIS 5-2.E & $I_E$ &  89.52 \\
3590-1-1 & 03:43:01 & VIS 4-2.H & $I_E$ & 560.52 \\
3590-1 & 03:53:13 & NISP DET21 & $J_E$ & 87.2448 \\
3590-1 & 03:55:27 & NISP DET21 & $H_E$ & 87.2448 \\
3590-1 & 03:57:41 & NISP DET21 & $Y_E$ & 87.2448 \\
3590-1-2 & 03:57:44 & VIS 4-2.H & $I_E$ &  89.52 \\
3590-2-1 & 04:00:47 & VIS 4-2.H & $I_E$ & 560.52 \\
3590-2 & 04:10:59 & NISP DET21 & $J_E$ & 87.2448 \\
3590-2 & 04:13:13 & NISP DET21 & $H_E$ & 87.2448 \\
3590-2 & 04:15:27 & NISP DET21 & $Y_E$ & 87.2448 \\
3590-3-1 & 04:18:33 & VIS 4-2.E,H & $I_E$ & 560.52 \\
3590-3 & 04:28:45 & NISP DET21 & $J_E$ & 87.2448 \\
3590-3 & 04:30:59 & NISP DET21 & $H_E$ & 87.2448 \\
3590-3 & 04:33:13 & NISP DET21 & $Y_E$ & 87.2448 \\
\enddata
\end{deluxetable}

\section{Empirical Point Spread Function Models}
\label{sec:psf}

Accurate stellar photometry with \dolphot{} requires a library of model PSFs for
each detector and filter. For example, PSFs for the \dolphot{} \jwst{} module \citep{Weisz:2024}
are provided by  WebbPSF (now called STPSF) \citep[e.g.,][]{Perrin2014}. No comparable, public resource exists for
\euclid{}. There is no tool to generate the models, and the PSF products available
from the \euclid{} repository are not suitable for precision resolved star photometry
\citep[e.g., the publicly released VIS PSF is not oversampled;][]{EuclidQ1}. The released models are also not uniformly available: the Q1 PSF file is missing for 10 of the 124 observations, including the observation of AM\nobreakdash-1.\footnote{\url{https://euclid.dataspace.esa.int/public-eds-help/Missing_PSF_files_for_some_observations.html}}

Empirical effective PSF (ePSF) models have been constructed for \euclid{} before, most notably by
\citet{Libralato:2024}, who, within the Euclid Consortium, built ePSFs for the VIS and NISP ERO imaging
of MW GC NGC\,6397. Our approach is similar in spirit, but aims to produce a general
purpose, publicly released library covering the full focal plane in every band, built
entirely from public data products.

For \dolphot{}, we construct the model \euclid{} PSFs empirically from the Q1 calibrated frames
themselves. While building empirical models is straightforward in principle, doing
so over the full focal plane is more involved in practice. Here, we describe our
procedure.

We build the models on the calibrated single-exposure (LE1) frames rather than the
stacked MER mosaics, so that the PSFs are sampled at the native pixel scale and in
the same distorted frame in which \dolphot{} performs photometry. 
Each VIS exposure
is read and stored per amplifier, such that a single pointing comprises
$36\times4 = 144$ quadrant frames of $\sim$2k$\times$2k pixels. Each NISP
exposure is taken through a single filter and comprises a $4\times4$ array of 16
detectors of $\sim$2k$\times$2k pixels.  For consistency with our photometry, which treats each VIS amplifier independently
(\S~\ref{sec:preprocessing}), we make no attempt to stitch the four quadrants
into a single image. We
therefore derive one PSF library per VIS quadrant and one per NISP detector and filter.

\subsection{PSF Star Selection}
\label{sec:psfstars}

We identify PSF stars by cross-matching each frame against \emph{Gaia} DR3
\citep{GaiaDR3}. We adopt \emph{Gaia} rather than an internal source finder because
it provides a clean, externally vetted list of \textit{bona fide} point sources, which is
particularly valuable given the sparseness of the high-latitude \euclid\ Q1 fields.

For each of the 124 \euclid{} observations covered by the Q1 release, we compute the sky footprint from the WCS, perform a cone search over that region, and propagate the \emph{Gaia} proper motions to the \euclid{} observation epoch before assigning each star pixel coordinates in every frame of that observation (i.e., 4 dithered exposures for NISP and 6 for VIS). We then retain only stars
that are (i) isolated, with no \emph{Gaia} neighbor inside a predefined PSF extraction region; (ii)
consistent with single, well-behaved point sources, requiring
\texttt{non\_single\_star}~$<0.1$, \texttt{classprob\_dsc\_combmod\_galaxy}~$<0.1$,
and \texttt{ruwe}~$<1.2$; and (iii) located more than half the PSF extraction region from the
frame edge. For each surviving star, we extract an image cutout over the PSF extraction region, which we define as a square $55\times55$ pixel window. This is a deliberately generous window ($5.5\arcsec$ for VIS, $16.5\arcsec$ for NISP) that
captures the extended PSF wings while leaving an outer border for local background
estimation. Over the same region we also extract the corresponding RMS array and data quality (DQ) extensions, and we associate the cutouts of a given \emph{Gaia} star across the different dithered exposures using the \texttt{photutils} \texttt{LinkedEPSFStar} data class.

We use the DQ arrays to select the usable pixels within each cutout, retaining only pixels whose first DQ bit is zero (i.e., not flagged as \texttt{INVALID}). We further discard outright any cutout that contains saturated pixels (flagged by bit 3 for VIS and bit 10 for NISP) or in which more than 25\% of the pixels are bad. Because we apply this rejection to each cutout individually, a star that saturates in the 561~s VIS exposures can still contribute through its 90~s exposures. In total, these cuts (bad pixels and/or saturation) exclude $\sim$20\% of the VIS PSF star candidates entirely (a median of 918 per VIS quadrant, and 920--1,016 for the three quadrants covering AM\nobreakdash-1). Because the Q1 footprint is sparse but vast, this selection still yields a large number of PSF stars: typically
$\sim$3,500 per VIS quadrant (i.e., per $\sim$2k$\times$2k single-amplifier readout) and $\sim$12,000 per NISP detector.

\subsection{Model Construction and Grid Packaging}

\begin{figure}[t]
\centering
\includegraphics[width=\columnwidth]{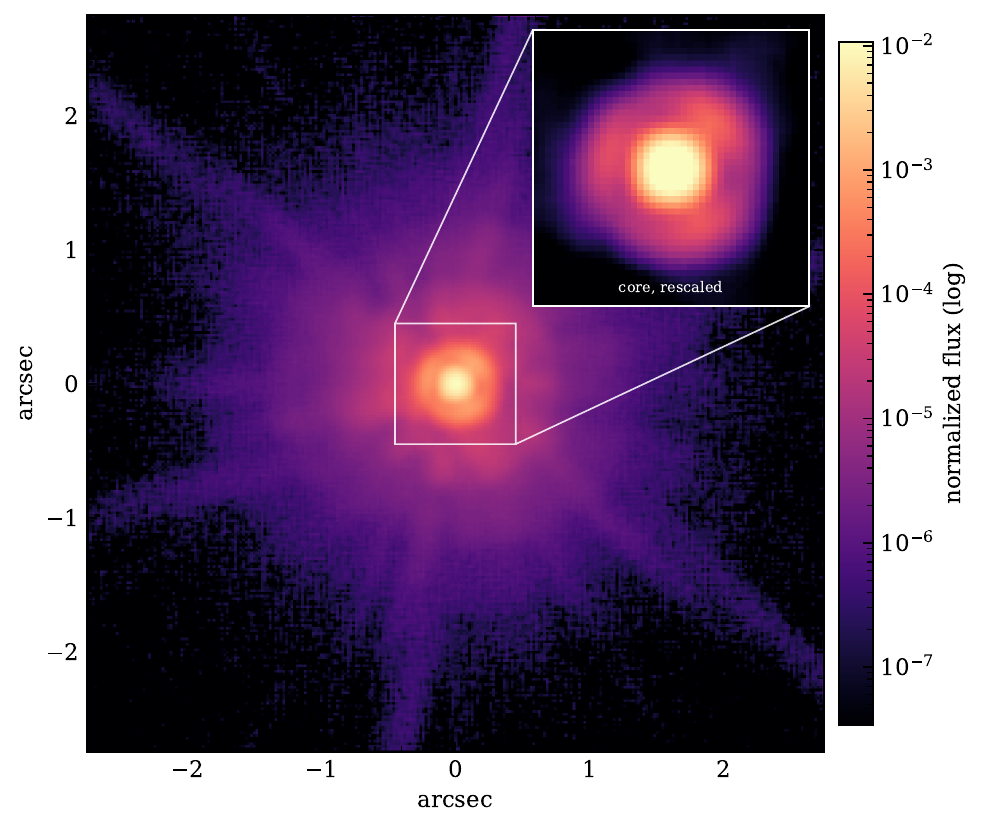}
\caption{An example of one of our empirical \euclid{} VIS PSFs for a representative detector quadrant
(DET34E), constructed from the Q1 calibrated frames with the iterative ePSF algorithm
(\S~\ref{sec:psf}). We show the central node of the $3\times3$ model grid, oversampled
by a factor of five relative to the native VIS pixel scale ($0.1\arcsec$), on a logarithmic stretch. The six diffraction spikes arise from \euclid's three-strut secondary mirror support. The inset shows the central $0.9\arcsec$ on a compressed logarithmic stretch, which reveals the three-lobed (trefoil) core.
\label{fig:psf_vis}}
\end{figure}

We build the empirical PSF for each detector from its full set of cutouts using the
iterative ePSF algorithm of \citet{Anderson2000}, as implemented in
\texttt{photutils} \citep{photutils}. We adopt a mode estimator when evaluating the ePSF model residuals, rather than the median estimator implemented in \texttt{photutils}. This choice is more computationally costly, but more robust against contaminants, such as imaging artifacts, background galaxies, or faint stellar neighbors missing from the \emph{Gaia} catalog. Because these contaminants only add flux, they skew the residual distribution to one side; the mode stays on its peak, i.e., the clean stars, while the mean and median are pulled into the tail. Locating that peak reliably requires a large PSF-star sample, which the Q1 footprint provides.

Prior to the fit, we subtract a local sky from
each cutout, estimating the background as the mode of the pixels in the outer border
of the cutout via a Gaussian kernel-density estimate. We construct each model
at an oversampling factor of 5 relative to the native pixels, spanning the full
$55\times55$ pixel cutout, and normalize it over that entire region (i.e., a normalization
radius of 55 pixels, equal to the cutout size) before iterating to convergence. 

Distortion is negligible across a single cutout. Instead, it matters when we place each star's cutouts from different dithers on a common sky coordinate system, which is mapped into each frame through that frame's WCS (via \texttt{LinkedEPSFStar}; \S~\ref{sec:psfstars}). The model PSFs therefore inherit the accuracy of the pipeline astrometry. Unlike \citet{Libralato:2024}, we do not solve jointly for distortion and PSF. Our tests in \S~\ref{sec:apcor_perf} suggest that any resulting broadening is small, though the larger DR1 datasets are needed to confirm this more thoroughly.

\begin{figure*}
\centering
\includegraphics[width=\textwidth]{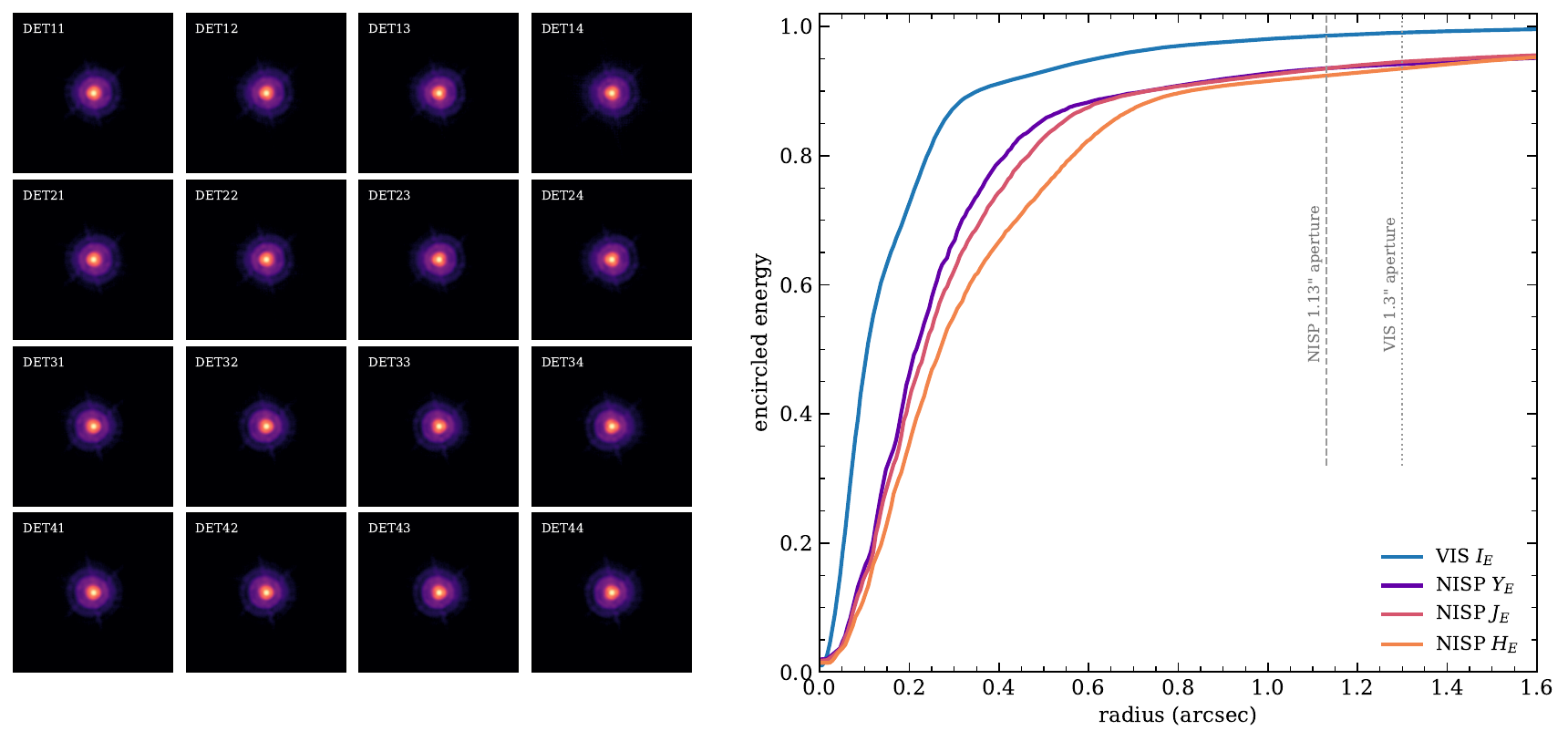}
\caption{Illustrative examples of our NISP PSF models and encircled energy (EE) curves. \textbf{Left:} Empirical NISP $H_E$ PSFs for all 16 NISP detectors, each the
central node of its $3\times3$ model grid, shown on a logarithmic stretch. The
detector-to-detector variation of the PSF is modest but clearly present.
\textbf{Right:} EE of the empirical VIS $I_E$ and NISP
$Y_E$, $J_E$, $H_E$ PSFs, normalized within the $55$-pixel model cutout. The dashed and
dotted vertical lines mark the NISP ($1.13\arcsec$) and VIS ($1.3\arcsec$, i.e., the
13-pixel aperture to which the VIS zero point is referenced) photometric apertures.
The VIS PSF encloses 99\% of its energy within $1.3\arcsec$, while the NISP PSFs
broaden with increasing wavelength.
\label{fig:psf_quality}}
\end{figure*}

Figures~\ref{fig:psf_vis} and \ref{fig:psf_quality} show the resulting models.
The VIS models recover the extended wings and the six diffraction spikes from \euclid's three-strut secondary mirror support, and a trefoil core with the same three-fold symmetry (Figure~\ref{fig:psf_vis}, inset). An azimuthal Fourier decomposition of the median-combined models over $0.4\arcsec$--$1.2\arcsec$ is dominated by the $m=3$ harmonic, indicating that this structure reflects the telescope optics rather than the model construction.

The left panel of Figure~\ref{fig:psf_quality} shows the
corresponding $H_E$ models for all 16 NISP detectors. The detector-to-detector
variation is modest but clearly present, which motivates our choice to derive an
independent library per detector rather than adopting a single focal plane average.

We package the models on a $3\times3$ spatial grid for each detector, matching the
grid format used by the WebbPSF/STPSF libraries in the \dolphot{} \jwst\ modules. At present we include only limited treatment of the PSF spatial variation across each chip. The NISP models are \emph{static}, i.e., the single detector-averaged ePSF is replicated across all nine grid points. For VIS, we exploit the fact that adjacent quadrants must share the same PSF at their common boundary. At each grid point on a chip edge or corner shared with other quadrants, we define the ePSF as the average of the \emph{static} ePSFs of all quadrants involved.

This is a coarse approximation of \euclid's true spatial PSF variation. A fully spatially varying model is a
natural next step, but a production-quality version will require the larger PSF star
samples expected with DR1. We therefore defer it to future work. Using ePSF grids from Consortium-internal reductions of the NGC\,6397 ERO imaging, \citet{Libralato:2024} find that the central pixel of the ePSF varies by a median of 1.2\% within a VIS quadrant and by medians of 1.6--2.6\% within a NISP detector. Our models capture little of this variation. However, solving for spatially variable residuals in \dolphot{} did not improve our AM\nobreakdash-1 photometry (\S~\ref{sec:overview}), suggesting that such variation is modest, or at least below what the current data can constrain.

\subsection{Aperture Corrections and Performance}
\label{sec:apcor_perf}

We determine aperture corrections from the same models, measuring the encircled
energy between the PSF fitted magnitude and a standard aperture.
For VIS, the zero point provided in the FITS header is calibrated to a 13-pixel circular aperture \citep{McCracken:2026}, so we measure aperture corrections to a circle of that radius.
For the NISP bands, the zero points provided in the FITS headers appear to be calibrated to the Q1 PSFs \citep{Polenta:2026}, which are normalized to unity in a square box of $2\arcsec \times 2\arcsec$.  We therefore adopt a standard aperture of radius $\sim 1.13\arcsec$ to match the area of that box.

As a quantitative check on the models, we compare their encircled energy (EE) with
published \euclid{} values. The right panel of Figure~\ref{fig:psf_quality} shows the
EE curves, measured as the cumulative azimuthally summed flux about the flux centroid. We find that the VIS PSF encloses 99\% of its
energy within the $1.3\arcsec$ (13-pixel) aperture to which the VIS zero point is
referenced. For NISP, we measure half-light radii of $r_{\rm EE50} = 0.22\arcsec,
0.23\arcsec, 0.27\arcsec$ and $r_{\rm EE80} = 0.42\arcsec, 0.47\arcsec, 0.57\arcsec$
in $Y_E, J_E, H_E$, broadening with wavelength as expected. This trend is broadly consistent with the independent ERO analysis of \citet{Annibali:2026},
who measure stacked-image FWHMs of $0.16\arcsec$ in $I_E$ and $0.47$--$0.49\arcsec$ in
$Y_E$, $J_E$, and $H_E$. Our NISP values sit comfortably
inside the $r_{\rm EE50} < 0.4\arcsec$ and $r_{\rm EE80} < 0.7\arcsec$ specification of
the released Q1 NISP PSF model \citep{Polenta:2026}.  Our EE measurements within $1\arcsec$
(91--98\%) are consistent with the $\sim$90\% values reported for both instruments by
\citet{Cuillandre:2025}. Our EE is normalized within the $55$-pixel
model cutout ($\pm2.75\arcsec$ for VIS and $\pm8.25\arcsec$ for NISP), so it does not
capture the faint, very extended wings that \citet{Cuillandre:2025} track to
$\sim$300\arcsec.
Our values are therefore slight overestimates. Normalizing to the full flux would lower the enclosed fraction at fixed radius by a few percent. This has little effect on $r_{\rm EE50}$ and $r_{\rm EE80}$, which are set by the core.

Applying the static VIS library to test frames, we find PSF
residuals that are small and comparable to those we obtain from the WebbPSF/STPSF models
for \jwst{} \citep{Weisz:2024}, and well within the range that \dolphot{} compensates for through its
on-the-fly PSF adjustments. 

The NISP models behave similarly, with one modest exception. Detector DET14 returns a slightly larger aperture correction than the other detectors. The size of this offset tracks the number of PSF stars used, not the presence of a few contaminating objects, suggesting that it reflects the statistics of the current PSF star sample rather than the detector itself. We expect the larger DR1 samples to resolve this issue. Regardless, the offset does not propagate into our photometry, because \dolphot{}'s aperture corrections measure the relationship between PSF fitting and aperture photometry within the standard aperture. Taken together, the EE comparison and the residual tests indicate that the empirical \euclid{} models are already accurate enough to support a first round of \dolphot{} photometry and validation.

\section{The \euclid{} \dolphot{} Module}
\label{sec:module}

\subsection{Overview}
\label{sec:overview}

The overall structure of a \dolphot{} module, including pixel masking, astrometric alignment, star finding, and catalog creation, is common to all supported cameras. \citet{Weisz:2024} describe each of these steps in detail for the \jwst{} modules. Here, we focus on what is new in version 2.1 and on the choices specific to \euclid{}.

All of our photometry uses \dolphot{} version 2.1, a recent upgrade to the widely used version 2.0 \citep{Dolphin:2016}. Several of the changes in this release were driven directly by \euclid{}.

The first is in the handling of PSF residuals. In version 2.0 and earlier, the PSF residual was assumed to be small, with the spatial variation of the PSF captured by the model itself. The Q1 data contain too few stars to measure a spatially variable PSF in this way, and version 2.1 therefore introduces supersampled (1x, 3x, or 5x), spatially variable PSF residuals.  

Our final photometry does not use spatially variable PSF residuals. The PSF variability appears minimal, and adding extra parameters to the PSF solution (effectively, measuring nine PSF residuals per image instead of one) did not improve the photometry.  We expect the spatially variable residuals to be more useful for data with many
more well-measured stars per frame, such as the crowded ERO fields, the Galactic
bulge imaging released with Q2, and the nearby galaxies expected with DR1, where the residual can be constrained at each grid point and the spatial variation of the
\euclid{} PSF is better sampled.  We will explore this issue in greater detail following DR1.

Version 2.1 introduces three new parameters, described fully in the \dolphot{} manual, that set the subpixel sampling, the
spatial resolution, and the smoothing of the PSF residual.  The first, \texttt{PSFresSuper}, sets
the subpixel sampling of the residual image and must be 1, 3, or 5 (3 by default). The second,
\texttt{PSFresSpatial}, sets the number of spatial nodes at which the residual is
evaluated across an image, with the residual computed at the image corners and at
evenly spaced points between them when it exceeds unity (1 by default, i.e., a single
residual per image). The third, \texttt{PSFresKernel}, sets the smoothing kernel applied to the
supersampled residual, either \texttt{fit} (a quadratic fit), \texttt{smooth} (a
boxcar smooth, the default), or \texttt{blend} (a mix of the two). In this analysis, we adopt
\texttt{PSFresKernel}$=$\texttt{blend} and leave \texttt{PSFresSpatial} at its default
(Table~\ref{tab_dolphot_params}), for the reasons given above.

The second major change, and the one that drove the version number, is to the star finding. Earlier versions of \dolphot{} were prone to identifying artifacts in the subtracted wings of bright stars as spurious detections on the second and later finding passes. We traced this to the noise model, which did not account for the uncertainty in the PSF model at locations where stars had been subtracted. Such false detections could survive in the photometric solution and, although they were ultimately rejected by the data-quality cuts, could contaminate the PSF adjustments and aperture corrections, corrupting the photometry by up to a few percent. This behavior is now resolved, and we find that \texttt{SecondPass} values greater than one are rarely beneficial.

A third, more visible change is more robust aperture corrections: the first two values printed to the console for each image should now be close to zero when photometry has run nominally.

As with version 2.0, version 2.1 has an OpenMP-parallel sibling that returns identical results (versions 3.0 and 3.1, respectively).
\subsection{Preprocessing}
\label{sec:preprocessing}

The \dolphot{} \euclid{} module provides separate preprocessing utilities for the VIS and NISP images. The two are similar enough that we describe them together here. As with our PSF models, the module does not reassemble the four quadrants of a VIS detector into a single image, since doing so would force the use of a single average noise for the whole chip rather than the correct noise for each quadrant.

The readout noise can be obtained in two ways: estimated from the RMS extension in comparison with the science image, or read directly from the FITS header. For VIS we find that the two agree almost always, and we therefore adopt the header \texttt{RDNOISE} value by default. For NISP, the header does not carry enough information to recover the effective readout noise once the multiple reads per group and groups per integration have been processed, and neither the documentation nor the help desk could supply this detail. There, we default instead to estimating the readout noise from the good pixels in the RMS and science images. In both cases, the utility can be switched to the alternative method with a user flag.

We then mask bad and saturated pixels using the data quality definitions of \citet{Polenta:2026} for NISP and \citet{McCracken:2026} for VIS, and we use the \texttt{SATURATE} keyword to flag additional saturated pixels in the VIS images.

Finally, we compute and apply pixel area maps from the Projection Value keywords, and we
derive third-order SIP coefficients for compatibility with the existing alignment routines
in \dolphot{}, which support SIP but not TPV-style distortion. No camera supported by \dolphot{}
prior to \euclid{} distributed its distortion solution in the TPV style. \dolphot{} uses the resulting SIP solution for the initial image registration and then fits residual offsets and a scale term for each image.

\section{Photometry of AM\nobreakdash-1}
\label{sec:phot}

We ran \dolphot{} on all NISP detectors and VIS quadrants overlapping an $80\arcsec\times80\arcsec$ box centered on the cluster (see Table~\ref{tab_euclid_exposures}). We adopted the VIS mosaic of the same region as the alignment reference. We ran \dolphot{} with its default parameters except for those listed in Table~\ref{tab_dolphot_params}. \citet{Weisz:2024} systematically explore how these parameters affect \jwst{} photometry.

Two of these choices follow directly from the changes described in
\S~\ref{sec:overview}. We adopt \texttt{NoiseMult}$=0.125$, well above the default of
$0.05$, because version 2.1 folds the contribution of bright neighbors' subtraction
residuals into the uncertainties through this parameter, which makes it both more
important and more sensitive to tune than in earlier versions. We also set
\texttt{PSFphotIt}$=2$, which iterates the PSF photometry solution and, in version 2.1,
substantially reduces the bias in the recovery of faint stars.

\begin{deluxetable}{rlrl}
\tablewidth{0pt}
\tablecaption{\dolphot{} parameters for AM\nobreakdash-1.  A detailed description of each parameter can be found in the DOLPHOT manual. \label{tab_dolphot_params}}
\tablehead{
\colhead{Parameter} &
\colhead{Value} &
\colhead{Parameter} &
\colhead{Value}}
\startdata
UseWCS & 2 & SigFind & 4.5 \\
Align & 2 & SigFinal & 5.0 \\
img\_raper & 3 & PSFphotIt & 2 \\
img\_rchi & 3 & FitSky & 2 \\
img\_rsky & 15 35 & SkipSky & 2 \\
img\_rsky2 & 4 7 & NoiseMult & 0.125 \\
img\_apsky & 20 35 & PSFresKernel & blend \\
img\_rpsf & 15 & CombineChi & 1 \\
MinSep & 1.5 & NISPvega & 0 \\
\enddata
\end{deluxetable}

To ensure that only \textit{bona fide} stars informed the PSF measurements and aperture corrections, we used an initial photometry pass to select high signal-to-noise ($SNR\ge10$ in every band and $\ge65$ overall), non-extended sources ($sharp\ge-0.04$). A second \dolphot{} run then used these stars for the PSF and aperture corrections to produce our final photometry.

The \dolphot{} catalog adopts the FITS keyword \texttt{MAGZEROP} for the VIS zero point, and either \texttt{ZPAB} or \texttt{ZPVEGA} for NISP, depending on the \texttt{NISPvega} setting. We set \texttt{NISPvega} to zero (Table~\ref{tab_dolphot_params}), so that our photometry is on the AB system.

\begin{figure*}
\centering
\includegraphics[width=\textwidth]{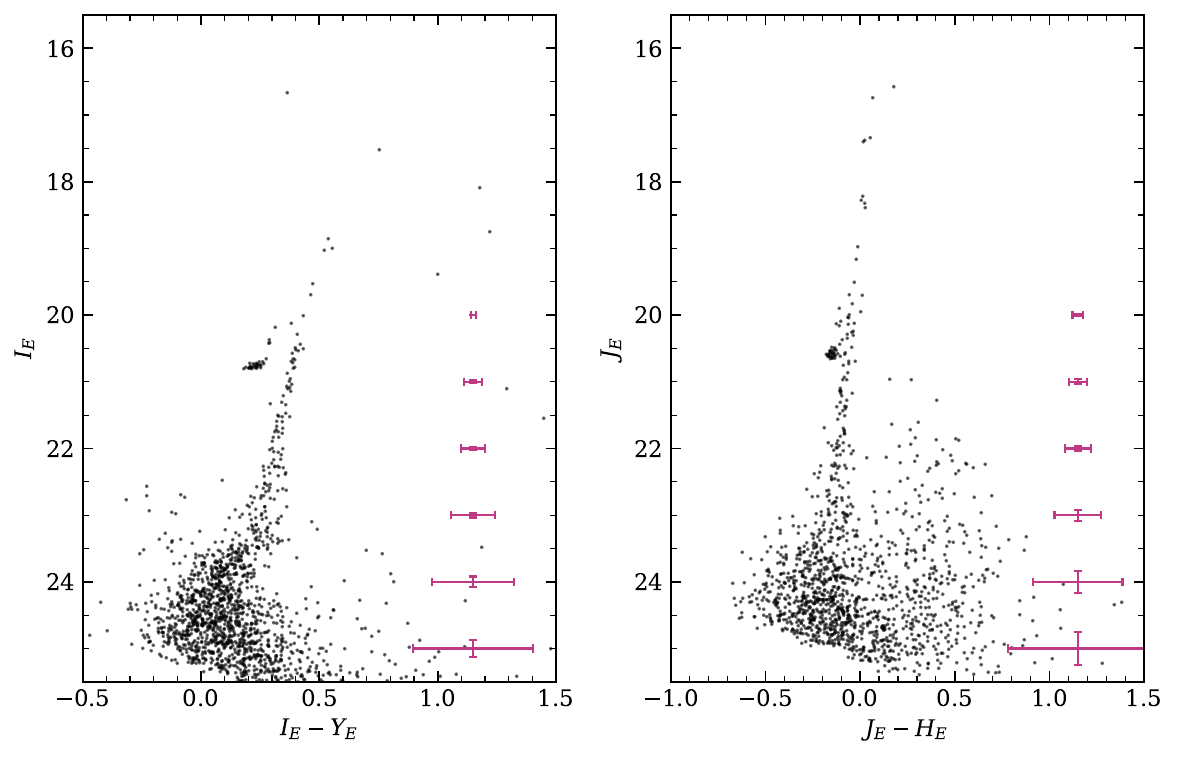}
\caption{\euclid{} CMDs of AM\nobreakdash-1 from our \dolphot{}
photometry, for all stars within a projected radius of $60\arcsec$ of the cluster center, which is larger than the $\sim1\arcmin$ square field shown in Figure~\ref{fig:am1_color}. \textbf{Left:} $I_E$
versus $I_E - Y_E$. \textbf{Right:} $J_E$ versus $J_E - H_E$. Error bars show the
representative photometric uncertainties measured from our artificial star tests (ASTs).
The main sequence (MS) is well populated to roughly a magnitude below the MSTO, and
the red giant branch (RGB) and red HB are clearly resolved.
\label{fig:euclid_cmd}}
\end{figure*}

We applied only signal-to-noise ($SNR\ge5$) and sharpness ($|sharp|\le0.2$) cuts to construct our final catalog. At the depth of the Q1 data, particularly the 87~s NISP exposures, any additional cuts substantially reduced the completeness on the main sequence. AM\nobreakdash-1 lies in the Euclid Deep Field South, which will be observed $\sim$40 times over the mission \citep{Mellier:2025}, compared with the single visit in Q1. These deeper VIS and NISP data should support more stringent cuts (e.g., to better separate stars from compact background galaxies, as in \citealt{Warfield:2023a}).
\citet{Weisz:2024} discuss the trade-offs among the \dolphot{} quality cuts in more detail.
Figure~\ref{fig:euclid_cmd} shows the resulting $I_E, Y_E$ and $J_E, H_E$ CMDs.  

\begin{figure*}
\centering
\includegraphics[width=\textwidth]{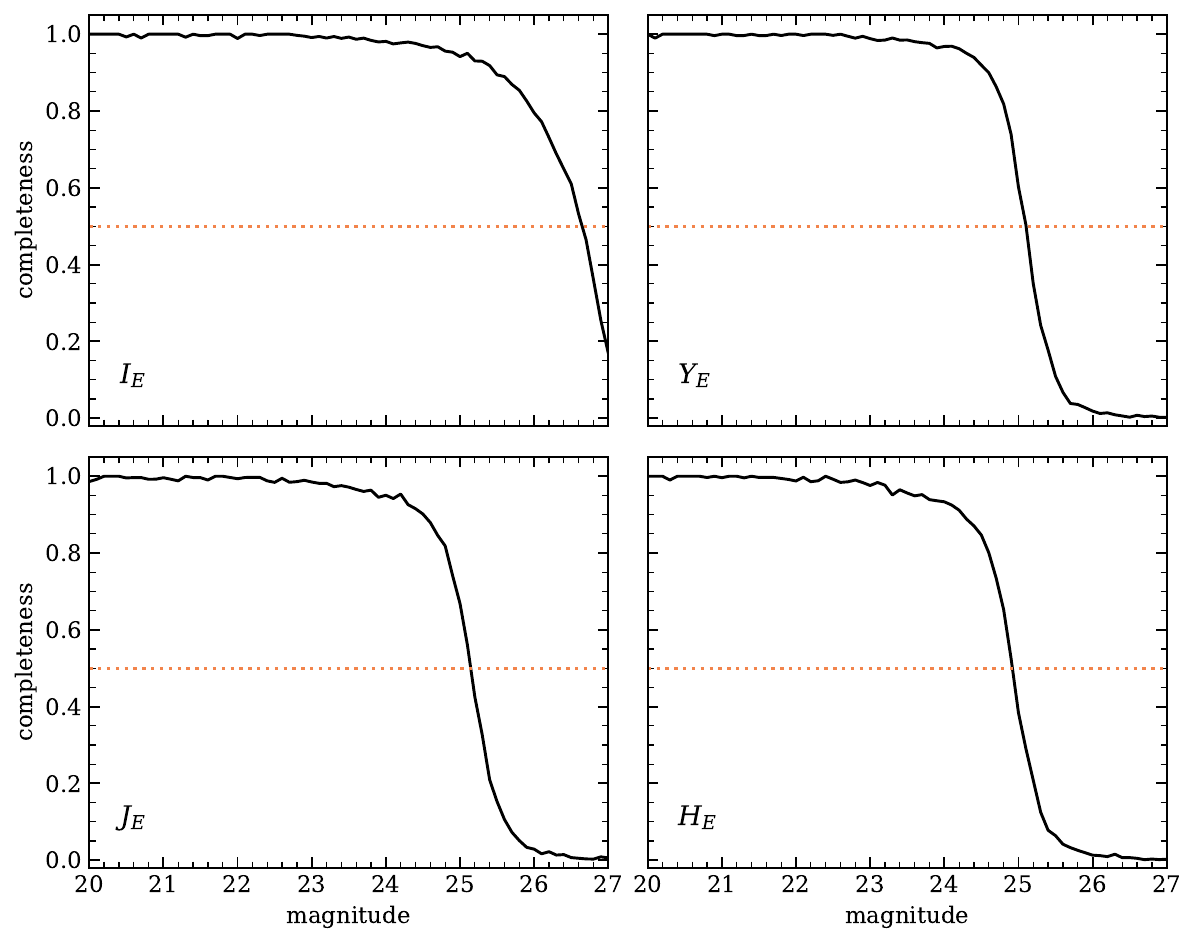}
\caption{Completeness as a function of input magnitude in each of the four
\euclid{} filters, measured from $\sim10^5$ artificial stars inserted into the images
and recovered with the same $SNR$ and sharpness cuts applied to our
science catalog. The orange dotted line marks 50\% completeness. We reach 50\%
completeness at $I_E = 26.6$, roughly 1.5~mag deeper than in the NIR bands
($Y_E = 25.1$, $J_E = 25.1$, $H_E = 24.9$). We do not probe to the completeness
limit in $I_E$, as it is not necessary for the current analysis.
\label{fig:completeness}}
\end{figure*}

To characterize the completeness and photometric uncertainties, we ran $\sim10^5$ ASTs. As always in \dolphot{}, the stars are added one at a time, and the photometry is performed in a small region around each inserted star and compared against the original catalog to determine whether the star is recovered, or whether it is blended with a nearby neighboring star. We apply the same $SNR$ and sharpness cuts to the recovered stars when computing the completeness.
We refer the reader to \citet{Weisz:2024} for a fuller description of \dolphot{} artificial star tests and their use in characterizing completeness and photometric bias.  We show the per filter completeness curves for AM\nobreakdash-1 in Figure~\ref{fig:completeness}.

\section{Discussion}
\label{sec:discussion}

\subsection{Comparison with Euclid Q1 Catalog}
\label{subsec:comp_euclid}

As a first test of our photometry, we compared it to the \euclid{} Q1 photometric catalog of \citet{Romelli:2026}. We accessed the catalog through the IRSA interface,\footnote{\url{https://caltech-ipac.github.io/irsa-tutorials/euclid-intro-mer-catalog/}} selecting point-like objects (\texttt{point\_like\_flag}~=~1 or
\texttt{point\_like\_prob}~$>$~0.7)
within a $0.1\times0.1$~degree box centered on the cluster.

We construct aperture (APHOT) and template-fitting (TPHOT) magnitudes
following the MER Photometry Cookbook\footnote{\url{https://euclid.esac.esa.int/dr/q1/dpdd/merdpd/merphotometrycookbook.html}},
adopting the 2FWHM aperture (i.e., a circular aperture of diameter twice the FWHM
of the worst-resolution band, measured on PSF-matched images). The reference
photometric measurement in the MER catalog is \texttt{FLUX\_DETECTION\_TOTAL},
the Kron flux \citep{Kron:1980} in the detection mosaic. For the VIS-detected sources considered
here (\texttt{VIS\_DET}~=~1) this is the total $I_E$ flux,
\begin{equation}
  F_{I_E} = F_{\rm det} \equiv \texttt{FLUX\_DETECTION\_TOTAL}.
\end{equation}
The MER catalog also provides a \texttt{FLUX\_VIS\_PSF} column, but
it is not used in the Cookbook's prescriptions for total fluxes, nor was its use recommended by the \euclid{} helpdesk. For the NISP
bands $b \in \{Y_E, J_E, H_E\}$, we obtain total fluxes by scaling the
detection flux by a color term measured between that band and VIS. For aperture
photometry,
\begin{equation}
  F^{\rm APHOT}_{b} = F_{\rm det} \,
  \frac{\texttt{FLUX\_\textit{b}\_2FWHM\_APER}}{\texttt{FLUX\_VIS\_2FWHM\_APER}},
\end{equation}
and for template-fitting photometry, where the color term is measured against the
VIS image smoothed to the resolution of band $b$,
\begin{equation}
  F^{\rm TPHOT}_{b} = F_{\rm det} \,
  \frac{\texttt{FLUX\_\textit{b}\_TEMPLFIT}}{\texttt{FLUX\_VIS\_TO\_\textit{b}\_TEMPLFIT}} .
\end{equation}
All MER fluxes are in $\mu$Jy, which we convert to AB magnitudes with
$m_{\rm AB} = -2.5\log_{10}(F/\mu{\rm Jy}) + 23.9$.

\begin{figure*}
\centering
\includegraphics[width=\textwidth]{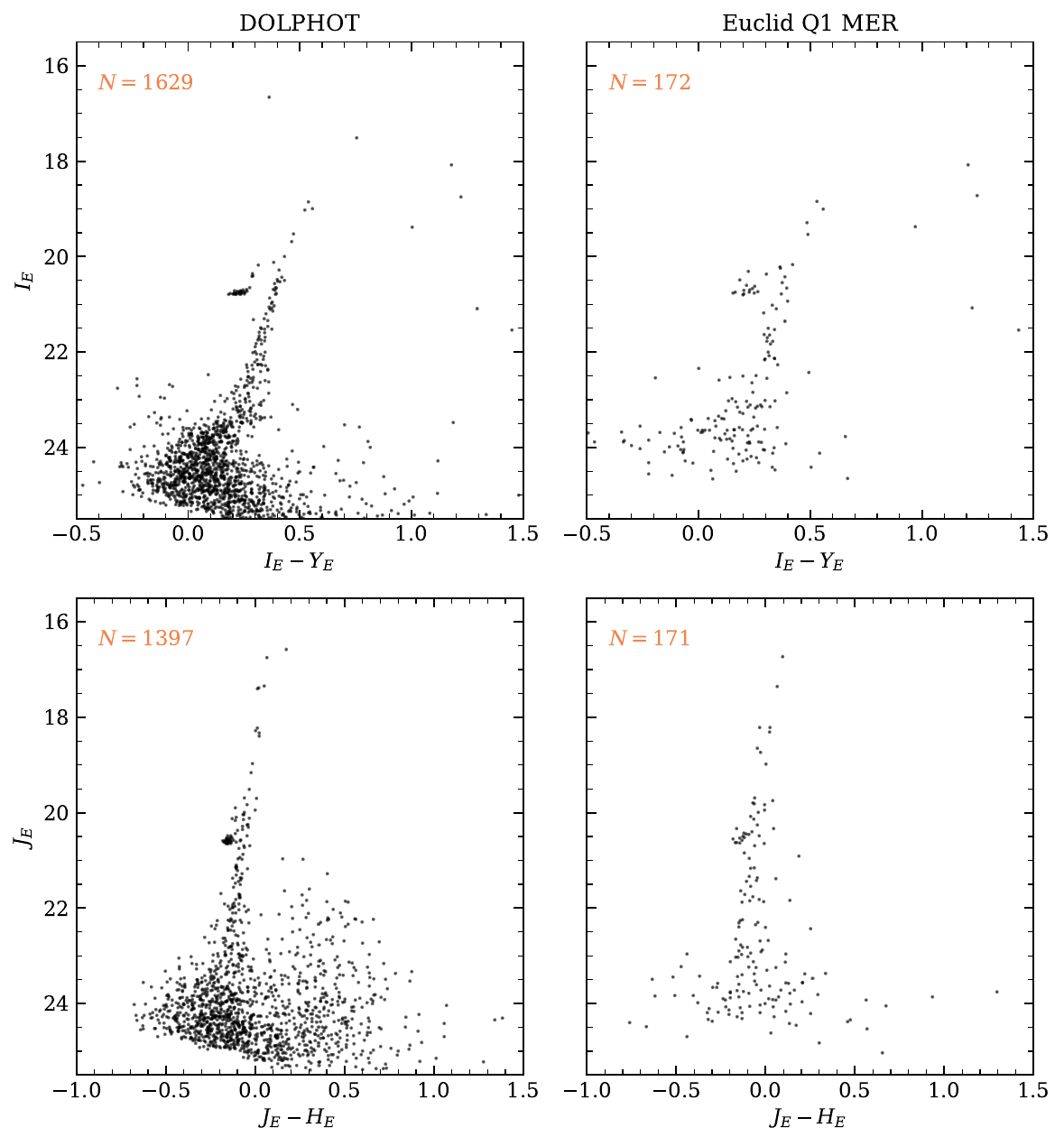}
\caption{A comparison of \euclid{} CMDs of AM\nobreakdash-1 from our \dolphot{} photometry (left) and from the
\euclid{} Q1 MER catalog (right), over the same $60\arcsec$-radius field and on identical
axes. MER photometry is APHOT in all bands as TPHOT is not available for VIS. Our catalog contains a factor of
$\sim$9 more sources over the same area and extends
to roughly a magnitude below the MSTO, whereas the MER catalog is essentially empty in the same region. 
\label{fig:dolphot_vs_mer_cmd}}
\end{figure*}

In Appendix~\ref{app:comparisons}, we show this comparison as a function of \dolphot{} magnitude. As noted by \citet{Romelli:2026}, the APHOT shows less scatter at bright magnitudes, while the TPHOT is tighter at the faint end. We find generally good agreement between our photometry and the MER catalog in all bands. In VIS, the offset (\dolphot{} minus MER) is $\Delta I_E=-0.003\pm0.002$~mag, after we remove sources fainter than $I_E=22$ and reject outliers. In NISP we compared against both flavors of MER photometry. The APHOT offsets are $\Delta Y_E=+0.012\pm0.003$, $\Delta J_E=+0.022\pm0.002$, and $\Delta H_E=+0.025\pm0.002$~mag. For the TPHOT comparison we did not impose the $I_E<22$ cut, since PSF fitting delivers reliable photometry to fainter magnitudes, and there we measured $\Delta Y_E=-0.044\pm0.003$, $\Delta J_E=-0.020\pm0.003$, and $\Delta H_E=-0.017\pm0.003$~mag.

The more consequential difference between the two catalogs is not their zero points but their depth. Figure~\ref{fig:dolphot_vs_mer_cmd} compares the CMDs built from our photometry and from the MER catalog over the same $60\arcsec$-radius field. Our catalog contains a factor of $\sim$9 more sources and populates the MS to roughly a magnitude below the MSTO, where the MER catalog is essentially empty.  AM\nobreakdash-1 is relatively uncrowded and we expect even greater contrast in more crowded regions, such as the ERO imaging, the Galactic bulge released as part of Q2, and new regions released with DR1. 

\subsection{Comparison with the VISTA Hemisphere Survey (VHS)}
\label{subsec:comp_vhs}

To assess the accuracy of our absolute photometric calibration, we compared our photometry against the VISTA Hemisphere Survey \citep[VHS;][]{McMahon:2013}, which provides an external reference independent of \euclid{}. The VHS uses the Two Micron All Sky Survey \citep[2MASS;][]{Skrutskie:2006} as its primary astrometric and baseline photometric calibrator: hundreds of 2MASS point sources act as standard stars in every VISTA Infrared Camera (VIRCAM) exposure, enabling a robust, continuous calibration of the broad-band $ZYJHK_s$ filters. We obtained VHS data within $10\arcmin$ of AM\nobreakdash-1 from the ESO Archive Science Portal; only $J$ and $H$ are available in this region. We converted the VHS Vega magnitudes to the AB system following \citet{GonzalezFernandez:2018}, and then to the \euclid{} photometric system using the Appendix~C conversions for normal stars of \citet{Schirmer:2022}.

We show this comparison in Appendix~\ref{app:comparisons}. We find excellent agreement in both bands: the mean offset (\dolphot{} minus VHS) is $\Delta J_E = +0.015 \pm 0.003$~mag and $\Delta H_E = -0.002 \pm 0.006$~mag. This agreement provides an external validation of our NISP photometry.

\subsection{Analysis of the AM\nobreakdash-1 Stellar Population}
\label{subsec:AM1_population}

As a final check on this first application of the \dolphot{} \euclid{} module, we compared our photometry against results from archival WFPC2 data CMD modeling. Using the WFPC2 imaging described in \S~\ref{hst_data}, we ran \dolphot{} exactly as for the \euclid{} data, except that we set \texttt{UseWCS}=1 to take advantage of the higher-order distortion corrections built into \dolphot{}.

Our aim here is to validate the \euclid{} photometry, not to redetermine the properties of AM\nobreakdash-1, which are already analyzed in detail by \citet{Dotter:2008b}. We therefore derive the stellar-population parameters from the well-characterized WFPC2 CMD and use them to test the \euclid{} CMDs. For a system newly discovered in \euclid{} imaging, with no existing \hst{} data, one would instead fit the \euclid{} CMDs directly to determine the age, metallicity, and distance.

\begin{figure*}
\centering
\includegraphics[width=0.49\textwidth]{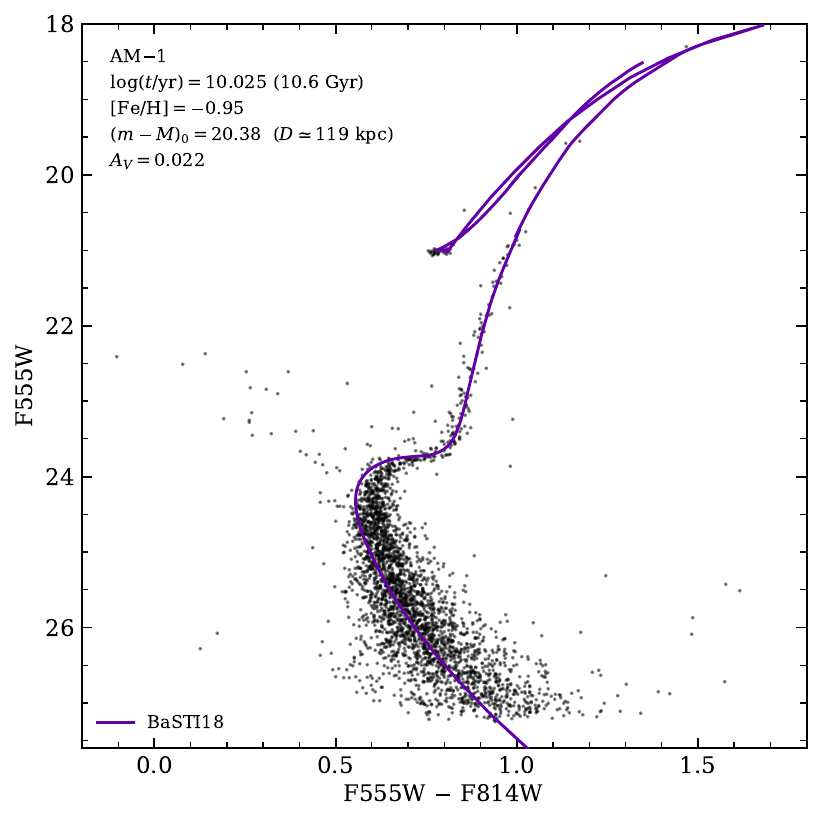}
\hfill
\includegraphics[width=0.49\textwidth]{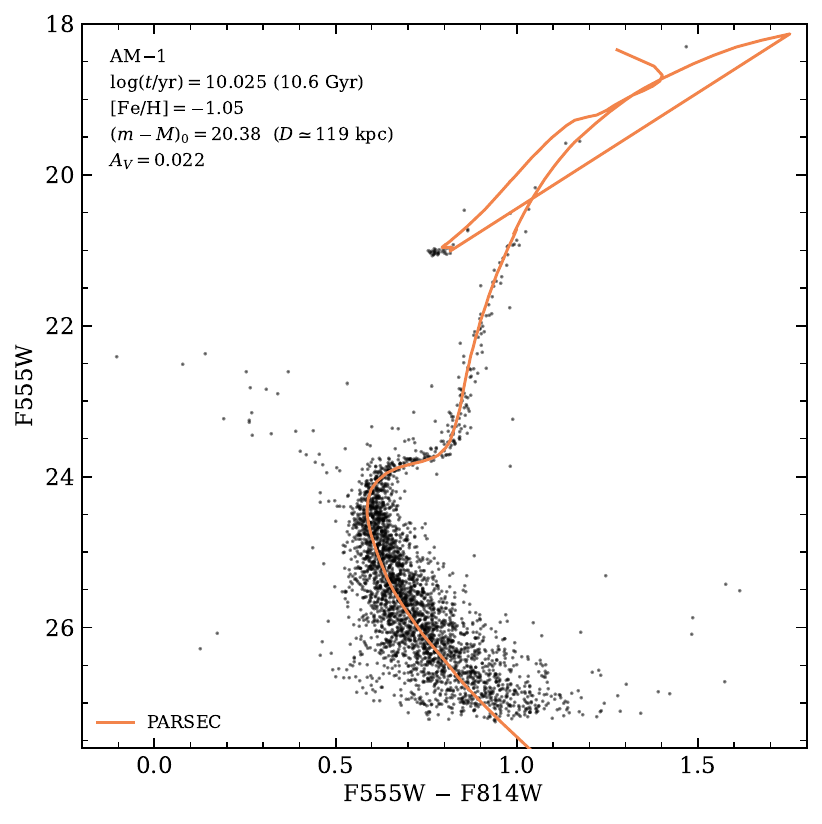}
\caption{WFPC2 color--magnitude diagram of AM\nobreakdash-1, with the best-fitting isochrone
overlaid: BaSTI18 in purple (left) and PARSEC in orange (right). We adopt
$\log(t/{\rm yr}) = 10.025$ (10.6 Gyr), $(m-M)_0 = 20.38$ ($D \simeq 119$ kpc), and
$A_V = 0.022$, with $[{\rm Fe/H}] = -0.95$ for BaSTI18 and $-1.05$ for PARSEC.
Both isochrones provide a good match to the MS, MSTO, and red giant
branch.
\label{fig:wfpc2_cmd}}
\end{figure*}

We applied the same goodness-of-fit cuts as for the \euclid{} data and used the CMD-fitting package MATCH \citep{Dolphin:2002} to determine the best-fitting distance and stellar population, adopting $A_V=0.022$ from the extinction maps of \citet{Schlafly:2011}. With the BaSTI (a Bag of Stellar Tracks and Isochrones) evolutionary models \citep[][]{Hidalgo:2018},\footnote{\url{http://basti-iac.oa-abruzzo.inaf.it}} we find a best fit for a pseudo-simple stellar population (SSP) with $\log(\mathrm{Age})=10.00$--$10.05$, $[\mathrm{Fe/H}]=-1.0$ to $-0.9$, and a distance modulus $\mu=20.38\pm0.01$. The isochrone plotted in Figure~\ref{fig:wfpc2_cmd} corresponds to the best fit for this pseudo-SSP.

We also repeated the fit with the PARSEC (PAdova and TRieste Stellar Evolution Code) models \citep{Bressan:2012}, for which we compute \euclid{}-system magnitudes using the YBC database of bolometric corrections \citep{Chen:2019}, recovering the same age and distance and a slightly lower metallicity, [Fe/H]$=-1.1$ to $-1.0$. We show both model sets throughout. Because MATCH evaluates the models on a discrete grid, these ranges reflect the grid spacing ($0.05$ in $\log t$ and $0.1$~dex in $[\mathrm{Fe/H}]$) rather than a measured width of the population. Combining the two model sets, we adopt $\log(t/{\rm yr}) = 10.025 \pm 0.025$ and $[\mathrm{Fe/H}] = -1.00 \pm 0.07$, where the metallicity uncertainty is the quadrature sum of the $0.05$~dex half-resolution and the $0.05$~dex systematic offset between the BaSTI18 and PARSEC central values.

Our WFPC2 fit reproduces the main features of AM\nobreakdash-1's stellar population identified by \citet{Dotter:2008b}. We recover an age ($\sim$10.6~Gyr) and a distance ($\mu=20.38$, $D\simeq119$~kpc) in good agreement with their determination, and, like them, we find an old population whose CMD is well described by a single isochrone and features a prominent red HB. The one parameter for which our fit differs meaningfully is the metallicity, which we discuss next.

Our best-fit metallicity, $[\mathrm{Fe/H}]\approx-1.0$, is more metal-rich than the
$[\mathrm{Fe/H}]=-1.5$ that \citet{Dotter:2008b} derive for AM\nobreakdash-1 from $\alpha$-enhanced
isochrone fits ($[\alpha/\mathrm{Fe}]=+0.2$). An older spectroscopic estimate,
$[\mathrm{Fe/H}]=-1.7$ from \citet{Suntzeff:1985}, is broadly consistent with the
\citet{Dotter:2008b} value but is based on only two red giants and is not a strong
constraint. However, part of the offset from \citet{Dotter:2008b} is a matter of
composition, as we use scaled-solar models. Converting the \citet{Dotter:2008b} metallicity onto the scaled-solar value using the relation from \citet{Salaris:1993} yields $[\mathrm{M/H}]\approx-1.4$, in better agreement with our value of $[\mathrm{M/H}]\sim-1.0$.  The remaining 0.4~dex difference is modest and generally within the range of metallicity differences in CMD fitting owing to different adopted physics and bolometric corrections in various stellar libraries \citep[e.g.,][]{Gallart:2005, Dolphin:2012, Weisz:2016, Skillman:2017}.  Spectroscopy is needed to pin down AM\nobreakdash-1's metallicity and detailed chemical composition.

\begin{figure*}
\centering
\includegraphics[width=\textwidth]{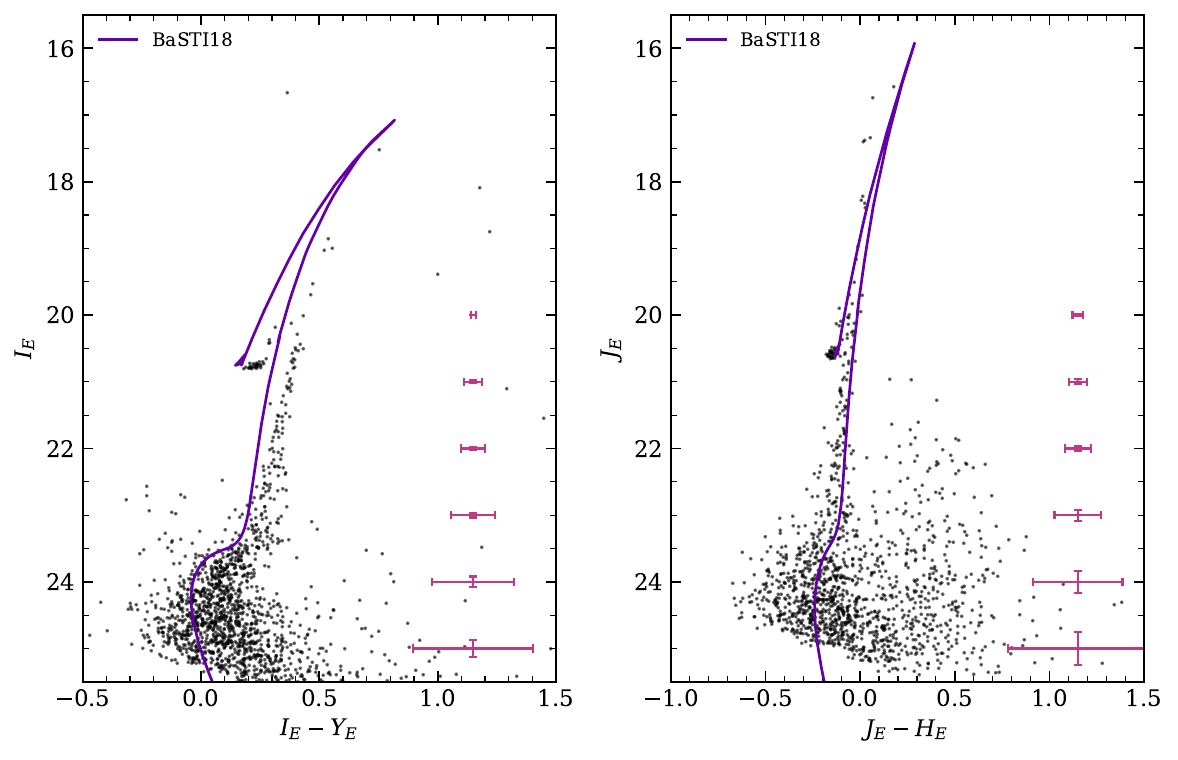}
\caption{Same as Figure~\ref{fig:euclid_cmd}, but with the BaSTI18 isochrone overlaid. We
adopt the parameters given in Figure \ref{fig:wfpc2_cmd}. The isochrone provides a reasonable match to both CMDs. However, on the $I_E - Y_E$ CMD it is systematically offset by $\sim0.07$~mag from the RGB, red HB, and MSTO. We discuss this offset
in \S~\ref{subsec:AM1_population} and Appendix~\ref{app:bc}.
\label{fig:euclid_cmd_iso_basti}}
\end{figure*}

\begin{figure*}
\centering
\includegraphics[width=\textwidth]{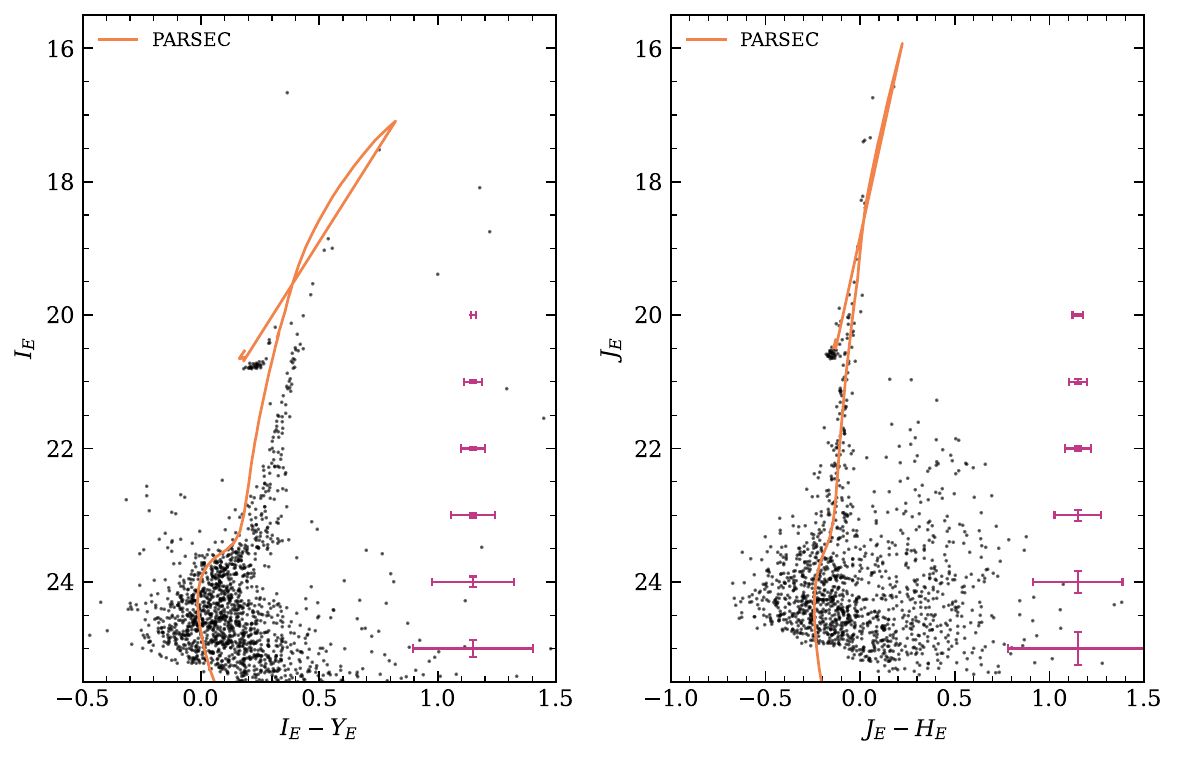}
\caption{Same as Figure~\ref{fig:euclid_cmd_iso_basti}, but for the PARSEC
models. The same $I_E - Y_E$ offset is present, indicating that it does not depend on the
choice of model set.
\label{fig:euclid_cmd_iso_parsec}}
\end{figure*}

We then overplot our best-fitting WFPC2 isochrone, in the \euclid{} photometric system, on the \euclid{} CMDs in Figures~\ref{fig:euclid_cmd_iso_basti} and \ref{fig:euclid_cmd_iso_parsec}. The isochrones match both the $I_E, Y_E$ and $J_E, H_E$ photometry reasonably well. The independent VHS comparison (\S~\ref{subsec:comp_vhs}), not the isochrone match, is our primary test of the $J_E$, $H_E$ absolute calibration. The small residual between the models and the observed upper RGB is comparable to that found in other model comparisons with optical and near-IR CMDs \citep[e.g.,][]{Dalcanton:2012a, Skillman:2017, McQuinn:2024a}.

However, in the $I_E-Y_E$ CMD, the isochrones are systematically offset from the data, most apparently in the color of the RGB. For $J_E$ and $H_E$ we can appeal to VHS (\S~\ref{subsec:comp_vhs}), but no comparable external reference is available for $I_E$ and $Y_E$. VHS covers only $J$ and $H$ in this field, and the Q1 MER catalog shares our \euclid{} zero points. We therefore use the WFPC2 photometry as an internal reference, predicting each star's $I_E$ and $Y_E$ from its WFPC2 color through quadratic transformations fit to the best-fitting isochrone. This is not a general purpose color transformation, but it should be adequate for this application and for this specific population.

Adopting $C \equiv F555W-F814W$, the quadratic fits to the isochrone colors are
\begin{equation}
    I_E-F814W = 0.303 + 0.305\times C - 0.027\times C^2,
\end{equation}
and
\begin{equation}
    Y_E-F814W = 0.889 - 0.759\times C + 0.129\times C^2.
\end{equation}

The equivalent fits using our best-fitting PARSEC isochrone are
\begin{equation}
    I_E-F814W = 0.255 + 0.304\times C - 0.021\times C^2,
\end{equation}
and
\begin{equation}
    Y_E-F814W = 0.767 - 0.613\times C + 0.070\times C^2.
\end{equation}

For each star we then predict $I_E$ and $Y_E$ from its measured $F555W$ and $F814W$ photometry, and compare those predictions with the \euclid{} magnitudes we measure with \dolphot{}. Because each prediction is referenced to the star's own measured $F814W$, the comparison is independent of distance, and it tests the two \euclid{} bands separately. Although extinction between the bands does not cancel, at the low reddening toward AM\nobreakdash-1 ($A_V=0.022$) it amounts to no more than a few mmag. As we show in Appendix~\ref{app:bc}, this comparison isolates the offset to $Y_E$.

\subsection{Structural Properties}
\label{subsec:structure}

The depth and areal coverage of the \euclid{} imaging also allow us to
characterize the spatial structure of AM\nobreakdash-1.  \citet{munoz:2018b} previously measured the structure of AM\nobreakdash-1, and comparing with their result further validates our photometry.

For the structural fit, we use only our \euclid{} \dolphot{} catalog, not the deeper WFPC2 imaging, which samples only the central few half-light radii.  The \euclid{} field covers the full radial extent of AM\nobreakdash-1 implied by the \citet{munoz:2018b} structural parameters.

We model the on-sky distribution of the member stars as an inhomogeneous Poisson point process and evaluate its unbinned likelihood, following \citet{Koposov:2015} and \citet{Weisz:2016}, to which we refer the reader for full details.

Briefly, the method fits the positions of the individual stars, rather than a binned surface density profile, modeling the cluster and a spatially uniform foreground/background component in a single likelihood. Fitting unbinned positions avoids the loss of spatial information and the sensitivity to bin placement inherent to binned fits, an advantage that is particularly valuable for a
sparse system such as AM\nobreakdash-1.

We restrict the fit to a clean sample of 306 candidate members, selected in the
range $19.5 < I_E < 24$ and within $100\arcsec$ of the cluster center
(Table~\ref{tab:am1}), the region over which the VIS and NISP coverage overlap completely. Beyond this radius the detector footprints only partially fill the field, so the fit's assumption of a uniform background over the full circular region would not hold. This selection isolates well-measured MS and RGB stars while limiting potential contamination from the noisy faint end. We fit three radial profiles commonly used for star clusters and
low-luminosity systems: an exponential, a  Plummer profile \citep{Plummer:1911}, and a
single-mass King model \citep{King:1966}. Each model shares a common centroid, half-light
radius ($r_h$), ellipticity, and position angle, together with a constant background
surface density. We sample the posterior with the affine-invariant ensemble sampler \texttt{emcee} \citep{ForemanMackey:2013} under uniform priors, so that the posterior is proportional to the likelihood, and quote posterior medians with 16th--84th percentile intervals.

We find that AM\nobreakdash-1 is well-described by all three profiles, with parameters that
 closely agree where they are directly comparable. The exponential and Plummer fits
return nearly identical circularized half-light radii, $r_h = 21.6\arcsec$ and
$20.8\arcsec$ respectively ($\simeq 12$~pc at $D = 119$~kpc), while the more
extended wings of the King model drive a modestly larger
$r_h = 27.0\arcsec$ ($15.7$~pc), as expected for that profile. The King fit
returns a low central concentration, $W_0 = 4.8^{+0.4}_{-0.5}$ and
$c \equiv \log(r_t/r_0) = 0.98 \pm 0.09$.

We find no evidence that AM\nobreakdash-1 is
elongated. The ellipticity is consistent with zero
($1 - b/a < 0.12$ at 90\% confidence). We therefore fix the ellipticity to zero and
do not report a position angle.

Our measurements are consistent with the ground-based structural parameters of
\citet{munoz:2018b}. They report an effective radius of
$16.5 \pm 1.1$~pc, in good agreement with our King value ($15.7$~pc) and modestly
larger than our exponential and Plummer values ($\simeq 12$~pc), as expected given
the differing assumed profiles. \citet{munoz:2018b} also find a round, regular morphology,
consistent with the circular shape we infer.

We combine these measurements with the stellar population fit of
\S~\ref{subsec:AM1_population} to estimate the present-day stellar mass and
mass-to-light ratio of AM\nobreakdash-1. The MATCH SSP fit returns the total mass formed, which
we convert to a standard Kroupa \citep{Kroupa:2001} mass range ($0.08$--$150\,M_\odot$) and reduce by the
mass lost to stellar evolution over $\sim$10.6~Gyr, retaining $\sim$60\% of the birth
mass as living stars and compact remnants \citep[e.g.,][]{LeitnerKravtsov:2011, Vincenzo:2016}. This yields a
present-day mass of $\sim$$2.1\times10^4\,M_\odot$. Combined with the luminosity
implied by our adopted $M_V = -5.1$ ($L_V \simeq 9.4\times10^3\,L_\odot$;
Table~\ref{tab:am1}), we obtain a $V$-band mass-to-light ratio
$M/L_V \simeq 2.2$. This is consistent with expectations for an old
GC ($M/L_V \sim 1.5$--2.5).

We have not searched for tidal tails around AM\nobreakdash-1, as such a search requires uniform photometry far outside the detectors analyzed here and is beyond the scope of this paper. A tidal-tail search is a natural application of the \dolphot{} \euclid{} module to the surrounding Q1 imaging and to future releases.

\subsection{AM\nobreakdash-1 in the Context of the Outer-Halo Globular Clusters}
\label{subsec:outerhalo}

AM\nobreakdash-1 appears to belong to a small
population of distant, diffuse, low-luminosity GCs in the outer MW
halo. Its size, luminosity, and distance are comparable to
other well-studied clusters of this class such as Pal\,3, Pal\,4, Pal\,14, and Eridanus \citep[e.g.,][]{Harris:2010, munoz:2018b}. Crater is
similarly extended ($r_h\sim20$~pc) and even more remote
\citep[$D\simeq145$~kpc;][]{Weisz:2016}.   These GCs have ages between $\sim$10 and $\sim$11~Gyr, except Crater, which is $\sim$7--7.5~Gyr old \citep[e.g.,][]{Stetson:1999, Dotter:2008b, Weisz:2016}.

\citet{Mackey:2005} identify these objects as
``young halo'' clusters and single out AM\nobreakdash-1 as
the most distant such cluster, a distinction later superseded by the discovery of Crater. AM\nobreakdash-1's dynamics point to an accreted origin. Using \emph{Gaia} orbits, \citet{Massari:2019} place
AM\nobreakdash-1 in the high-energy group of clusters that appear to have been accreted from
low-mass progenitors that cannot (yet) be individually identified. The chemodynamical
analysis of \citet{Callingham:2022} classifies AM\nobreakdash-1 as accreted but
unassociated with any known accretion event, grouping it with Crater, Pal\,3, Pal\,4,
Pal\,14, and Eridanus.  

More broadly, our study of AM\nobreakdash-1 illustrates the promise of \euclid{} for the study of faint stellar systems throughout the MW halo. \euclid{} unites two capabilities that have rarely been available in a single dataset: the angular resolution and stable PSF needed to resolve individual stars well below the MSTO, and the wide-area coverage needed to survey large volumes of the halo uniformly. Resolved star studies of systems like AM\nobreakdash-1 have traditionally relied on targeted \hst{} or \jwst{} imaging of one object at a time, whereas wide ground-based surveys supply the area but not the angular resolution to cleanly separate members from contaminants. 

The \euclid{} Wide Survey footprint will encompass $\sim100$ known Galactic GCs, including distant outer-halo clusters such as Pal\,4 and NGC~2419, a few dozen UFDs, and nearly two hundred galaxies within 5~Mpc \citep[e.g.,][]{Pace:2025a}. DR1 should include several southern UFDs (e.g., the Grus and Tucana systems) and dozens of nearby galaxies such as those of the Sculptor group. Deep ground-based imaging is also advancing rapidly. Rubin Observatory commissioning data have already quadrupled the sample of candidate member stars in the ultra-faint satellite Virgo~III \citep{Pai:2026}, while \euclid{} adds the space-based angular resolution and near-infrared coverage needed to push further into the crowded, faint regime. We anticipate that the public \dolphot{} \euclid{} module will support uniform resolved star studies of MW halo satellites and outer halo GCs as the survey grows.

\section{Summary}
\label{sec:summary}

We have presented a \euclid{} study of AM\nobreakdash-1, the only Galactic GC in the \euclid{} Q1
release and one of the most distant, diffuse GCs known in the MW. To measure its
resolved stars, we developed a \euclid-specific module
for \dolphot{} that builds empirical PSF models directly from the public Q1 calibrated
frames. Our main results are:

\begin{itemize}

\item Our \dolphot{} photometry recovers the MS of AM\nobreakdash-1 to roughly a
magnitude below its MSTO, with $5\sigma$ point-source depths of $I_E = 27.1$,
$Y_E = 25.3$, $J_E = 25.5$, and $H_E = 25.3$ (Table~\ref{tab:am1}), $\sim$1--1.5~mag deeper than the public Q1 data set
\citep[$26.0$, $23.8$, $24.0$, $24.0$ in $I_E, Y_E, J_E, H_E$;][]{Romelli:2026}.

\item Our WFPC2 and \euclid{} CMDs are consistent with an old, intermediate-metallicity stellar
population, with an age in agreement with the earlier analysis of \citet{Dotter:2008b}. The
best-fitting isochrone reproduces both \euclid{} CMDs reasonably well, but is systematically
offset from the observed $I_E - Y_E$ sequence. We trace this $\sim0.07$~mag offset to the $Y_E$
band, although we cannot determine whether it originates in the $Y_E$ zero point or
in a common error shared by the available bolometric corrections.

\item Our photometry agrees with the public MER catalog to $\lesssim0.05$~mag in
each available band, where the catalogs overlap. Our NISP photometry agrees with the independently calibrated VHS
catalog to $\Delta J_E = +0.015 \pm 0.003$ and $\Delta H_E = -0.002 \pm 0.006$~mag,
providing an external check on our absolute calibration.

\item From our \euclid{} photometry, we measure $r_h\sim$12--16~pc for AM\nobreakdash-1, which
places it among the most physically extended GCs known in the MW. The
cluster has a low central concentration and shows no evidence for
elongation. These parameters are consistent with the previous
ground-based measurements of \citet{munoz:2018b}. AM\nobreakdash-1's size, luminosity, and
orbit place it among the distant outer halo clusters that dynamical studies identify
as accreted \citep{Massari:2019, Callingham:2022}.

\item We present \dolphot{} v2.1, which includes several \euclid{}-motivated updates that apply to every camera \dolphot{} supports. These updates include a revised noise model that prevents residuals in the subtracted wings of bright stars from producing spurious detections. Version 2.1 also introduces supersampled, spatially variable PSF residuals, and more robust aperture corrections. OpenMP-parallel builds (v3.0 and v3.1) return results identical to v2.0 and v2.1.

\end{itemize}

Taken together, these results demonstrate that, even in the absence of a full public release of detailed calibration products, \euclid{} can deliver precise, deep,
resolved star photometry of low-luminosity stellar systems in the MW halo, well beyond what the public catalogs currently provide. As part of our \euclid{} Guest
Investigator Program (24-EGIP24-0005), we are publicly releasing a first version of the \dolphot{}
\euclid{} module so that the broader community can carry out similar analyses using DR1 data.

\begin{acknowledgments}
We thank the referee, Luigi Bedin, for a thoughtful and constructive referee report that improved the paper. This material is based upon work supported by the National Aeronautics and Space Administration through the Euclid Guest Investigator Program (ROSES-2024 D.18) under award No.\ 80NSSC25K7734 (proposal 24-EGIP24-0005; PI A.\ Savino).

This research has made use of the NASA/IPAC Infrared Science Archive, which is funded by the National Aeronautics and Space Administration and operated by the California Institute of Technology. We thank the IRSA Helpdesk for their assistance with the \euclid{} Q1 data retrieval and for clarifying the availability of the VIS exposures of AM\nobreakdash-1.

Based on observations made with the NASA/ESA Hubble Space Telescope, obtained from the Mikulski Archive for Space Telescopes (MAST) at the Space Telescope Science Institute (STScI). STScI is operated by the Association of Universities for Research in Astronomy, Inc., under NASA contract NAS5‑03127.

This research has made use of the SVO Filter Profile Service ``Carlos Rodrigo'', funded by MCIN/AEI/10.13039/501100011033/ through grant PID2023-146210NB-I00.

This paper was constructed and written by the authors, but portions were edited with the assistance of Claude Opus v4.8 and v5.0, large language models developed by Anthropic, to improve readability in June 2026.  Claude Opus v5.5 was used for a final proofread (i.e., to catch typos, improve clarity, check for redundancy and repetition) in October 2026. All content was reviewed and verified by the authors.

\end{acknowledgments}

\section*{Data Availability}

All of the imaging analyzed in this paper is public. The \euclid{} VIS and NISP
observations of AM\nobreakdash-1 are part of \euclid{} Quick Data Release~1 and were
retrieved from the NASA/IPAC Infrared Science Archive (Observation~ID~3590). The archival \hst/WFPC2 observations
(GO-6512) were retrieved from the Mikulski Archive for Space Telescopes. The VISTA
Hemisphere Survey photometry was obtained from the ESO Archive Science Portal, and the
\emph{Gaia} DR3 astrometry used for PSF star selection from the ESA \emph{Gaia} Archive.

The \dolphot{} source code, the \euclid{} module, and the empirical PSF libraries
described in \S~\ref{sec:psf} are distributed from the \dolphot{} website
(\url{http://americano.dolphinsim.com/dolphot/}). The
\dolphot{} photometric catalog of AM\nobreakdash-1 will be made available upon request.

\facilities{Euclid (VIS, NISP), HST (WFPC2), VISTA (VIRCAM), Gaia}

\software{\dolphot{} \citep{Dolphin:2016}, MATCH \citep{Dolphin:2002}, emcee \citep{ForemanMackey:2013},
\texttt{photutils} \citep{photutils}, Astropy \citep{astropy:2013, astropy:2018, astropy:2022},
NumPy \citep{numpy}, SciPy \citep{scipy}, Matplotlib \citep{matplotlib},
astroquery \citep{astroquery}}

\clearpage
\appendix

\section{Photometric Comparisons and Residuals}

\subsection{Catalog Comparison Residuals}
\label{app:comparisons}

\begin{figure*}
\centering
\includegraphics[width=0.82\textwidth]{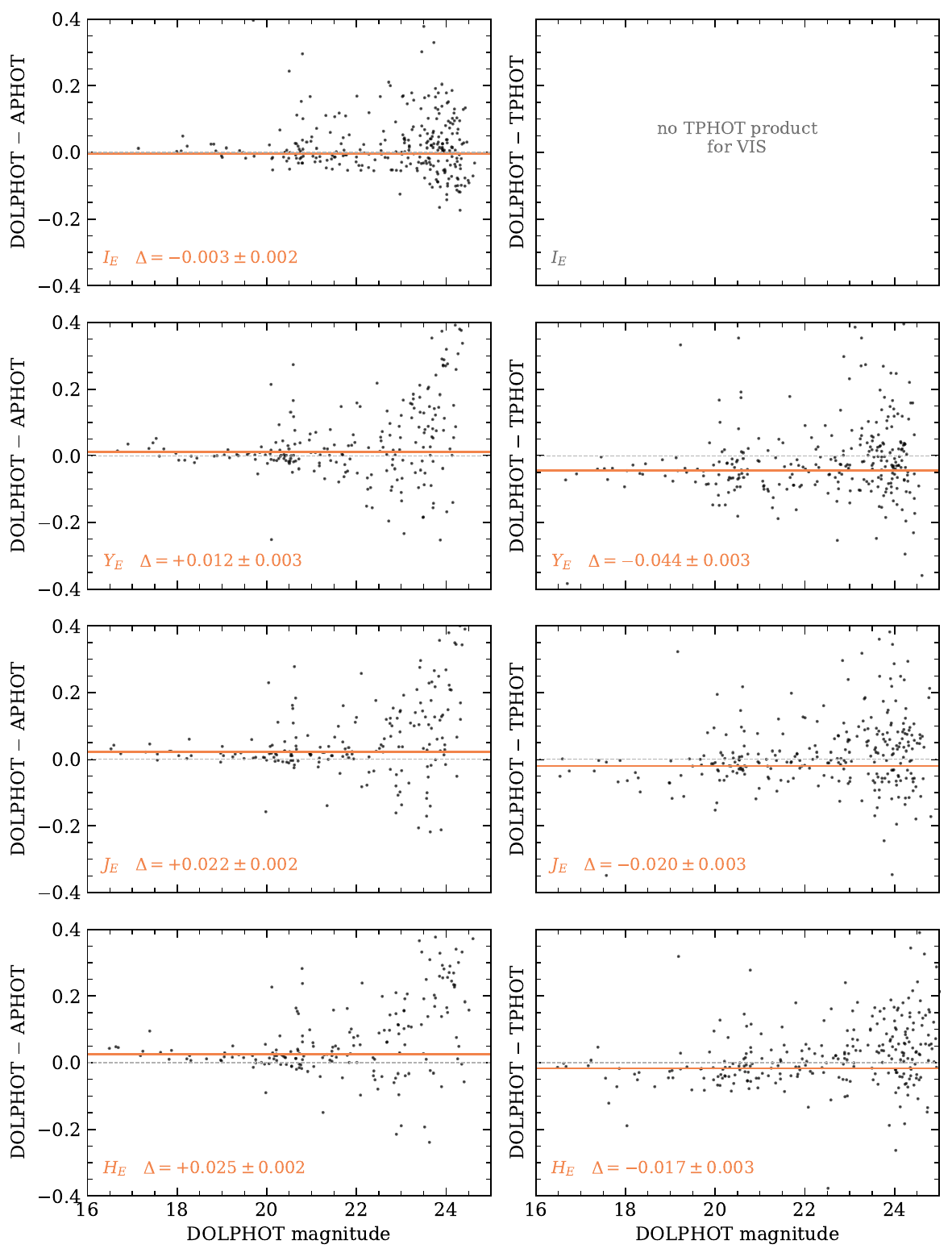}
\caption{Difference between our \dolphot{} photometry and the \euclid{} Q1 MER
catalog as a function of \dolphot{} magnitude, for the MER aperture photometry
(APHOT, left) and template-fitting photometry (TPHOT, right). The orange line marks
the offset given in \S~\ref{subsec:comp_euclid}, listed in the lower right of
each panel. We find generally good agreement, with all offsets $\le 0.05$ mag. The
MER catalog provides no template-fitting measurement in VIS, and so the
corresponding panel is left blank. We caution that the APHOT residuals show a
pronounced one-sided tail toward brighter \dolphot{} magnitudes, which grows toward
longer wavelengths and is consistent with a fixed aperture absorbing flux from
neighboring stars as the PSF broadens; the quoted offsets are therefore measured
after rejecting outliers.
\label{fig:q1_comparison}}
\end{figure*}

\begin{figure*}
\centering
\includegraphics[width=\textwidth]{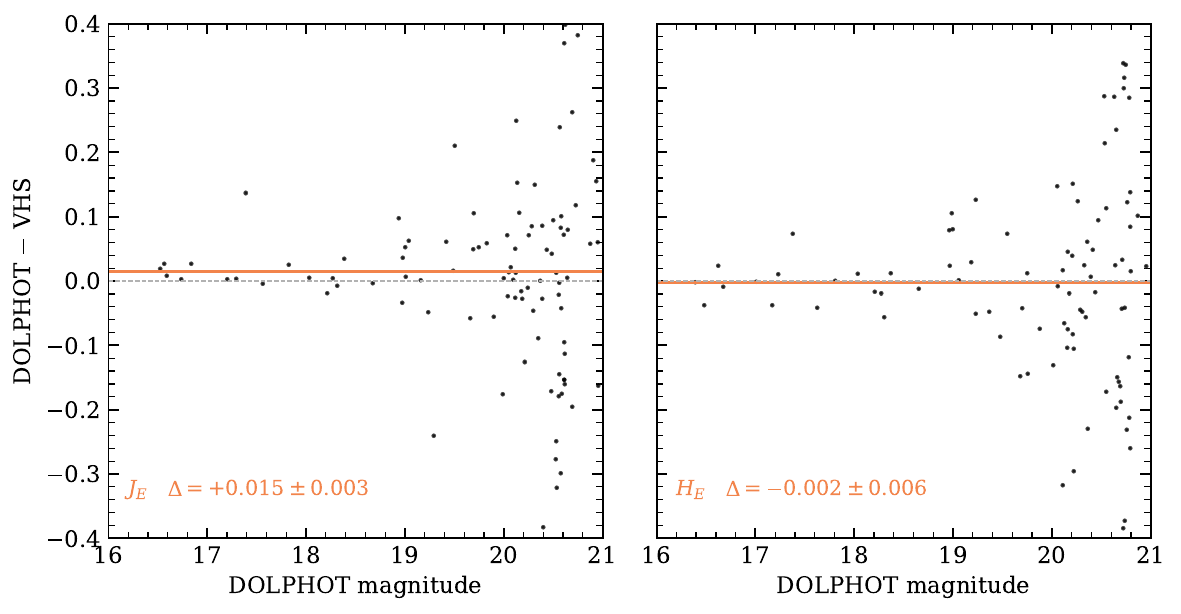}
\caption{Difference between our \dolphot{} photometry and the VHS catalog,
converted to the \euclid{} photometric system, as a function of \dolphot{}
magnitude. We restrict the comparison to $16 < m < 21$, where VHS is complete. The
orange line marks the mean offset. We find excellent agreement in both bands
($\Delta J_E = +0.015 \pm 0.003$, $\Delta H_E = -0.002 \pm 0.006$). Because VHS is
calibrated independently of \euclid{}, this comparison provides an external check
on our photometry.
\label{fig:vhs_comparison}}
\end{figure*}

Figures~\ref{fig:q1_comparison} and \ref{fig:vhs_comparison} show the photometric comparisons between \dolphot{} and external catalogs discussed in
\S\S~\ref{subsec:comp_euclid} and \ref{subsec:comp_vhs}. All offsets quoted, in both the main text and these figures, are uncertainty-weighted means measured after outlier rejection.

\subsection{Bolometric Correction Residuals}
\label{app:bc}

\begin{figure*}
\centering
\includegraphics[width=\textwidth]{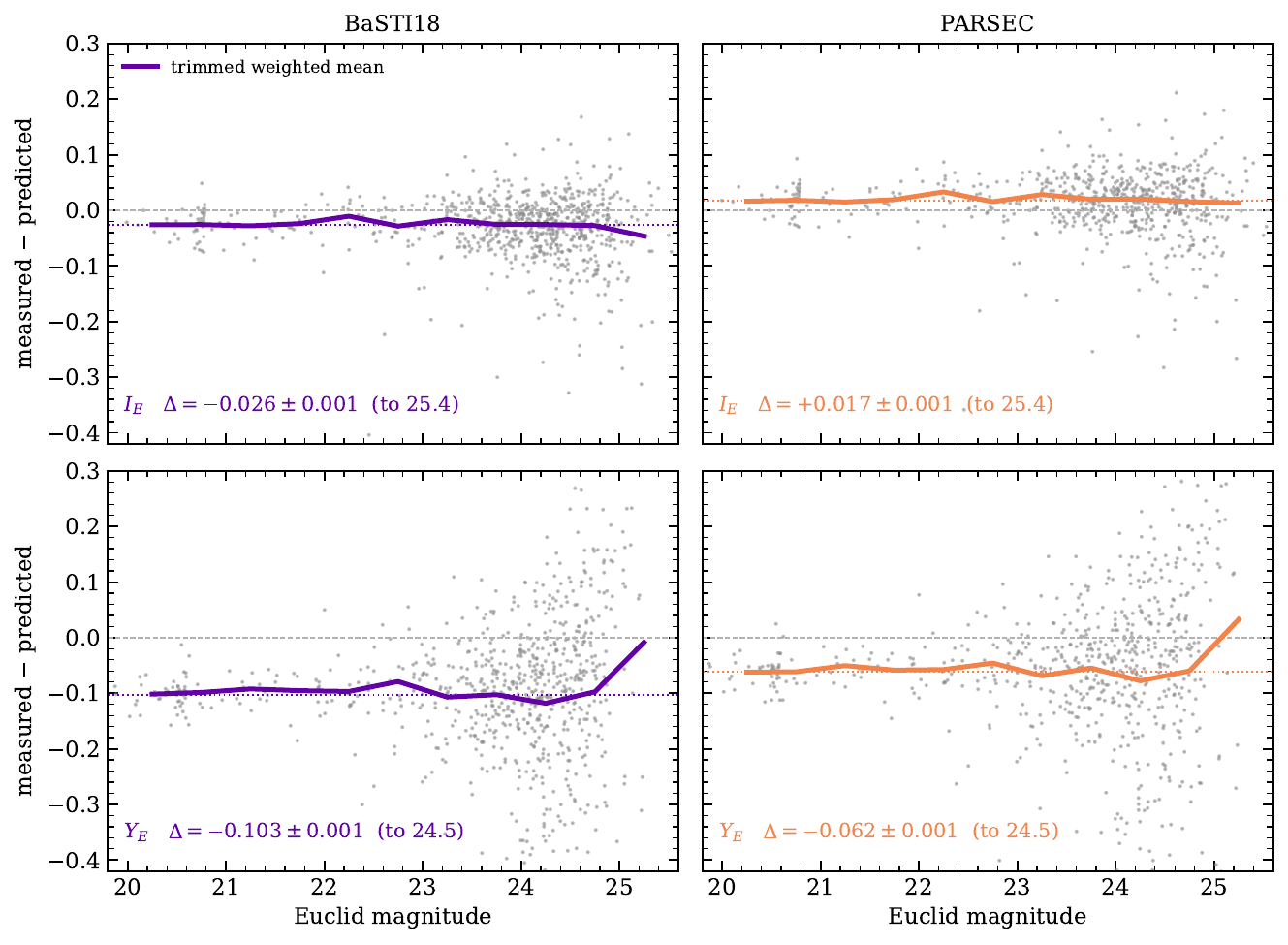}
\caption{Measured \euclid{} photometry minus that predicted from the WFPC2
$F555W, F814W$ photometry through the bolometric corrections, as a function of
magnitude, for BaSTI18 (left) and PARSEC (right). Because $F814W$ cancels from
both sides, this comparison is independent of distance and, at the low reddening toward AM\nobreakdash-1, effectively unaffected by extinction. 
Colored lines
show the running offset, and the dotted line and label give the offset over the
full range, measured to each band's 90\% completeness limit. The measured $I_E$
falls between the two sets of predictions ($-0.026$ and $+0.017$). In contrast, the measured $Y_E$ is brighter than predicted by both sets
of corrections ($-0.103$ and $-0.062$). Taken at face value, this suggests that the
$Y_E$ zero point, or the $Y_E$ system response adopted in the bolometric
corrections, is in error.
\label{fig:bc_residuals}}
\end{figure*}

We provide more detail on which of the two \euclid{} bands is responsible for the $I_E - Y_E$
color offset seen in Figures \ref{fig:euclid_cmd_iso_basti} and
\ref{fig:euclid_cmd_iso_parsec}. For each star, we predict its \euclid{} magnitude
from its WFPC2 $F555W, F814W$ photometry through each set of bolometric
corrections, and compare with the measured value down to the 90\% completeness limit of the \euclid{} data ($I_E < 25.5$, $Y_E < 24.5$). Figure~\ref{fig:bc_residuals} shows the resulting residuals as a function of magnitude for each set of bolometric corrections.

The measured $I_E$ falls between the two isochrone sets ($-0.026$ with BaSTI18,
$+0.017$ with PARSEC), and for either set the offset is smaller than the
$\sim$0.043~mag disagreement between the model-predicted \euclid{} magnitudes
themselves. The $I_E$ offset is therefore unlikely to be the cause of the observed CMD offset.
For $Y_E$, in contrast, both model sets predict magnitudes 0.06--0.10~mag fainter
than we measure. The residuals are flat with magnitude and color, i.e., the offset
appears to be constant, which is consistent with an error in the $Y_E$ zero point or
in the system response shared by the \euclid{} bolometric corrections adopted by the
BaSTI and PARSEC models. Filter width alone is unlikely to explain this offset. $I_E$ is the broadest \euclid{} filter in fractional width (FWHM/$\lambda \approx 0.55$), yet it agrees with the models, whereas $Y_E$ (FWHM/$\lambda \approx 0.24$) is comparable to the \jwst/NIRCam wide filters \citep{Rodrigo:2012, Rodrigo:2020, Rodrigo:2024}. The publicly available NISP passbands are pre-launch determinations \citep{Schirmer:2022}, so in-flight characterization of the $Y_E$ response may help resolve this offset.

\clearpage
\bibliographystyle{aasjournalv7}
\bibliography{citations}

\end{document}